\documentclass[showpacs,twocolumn,prb,eqsecnum]{revtex4}
\usepackage{amsmath}
\usepackage{epsfig,color}
\usepackage{graphicx}
\usepackage{dcolumn}
\usepackage{bm}
\newcommand{\nn}{\nonumber}
\newcommand{\beq}{\begin{equation}}
\newcommand{\eeq}{\end{equation}}
\newcommand{\bea}{\begin{eqnarray}}
\newcommand{\eea}{\end{eqnarray}}
\newcommand{\bwt}{\begin{widetext}}
\newcommand{\ewt}{\end{widetext}}

\newcommand{\vf}{v_{\mathrm{F}}}
\newcommand{\kf}{k_{\mathrm{F}}}
\newcommand{\ef}{E_{\mathrm{F}}}
\newcommand{\bkf}{{\bf k}_{\mathrm{F}}}

\newcommand{\G}{{\bar\Gamma}^{\Omega}_{\alpha\beta;\gamma\delta}}
\newcommand{\GRPA}{\Gamma^{\Omega, \mathrm{RPA}}_{\alpha\beta;\gamma\delta}}
\newcommand{\Gos}{\Gamma^\Omega_{\alpha\beta;\gamma\delta}}
\newcommand{\Gqs}{\Gamma^q_{\alpha\beta;\gamma\delta}}
\newcommand{\bk}{{\bf k}}
\newcommand{\bp}{{\bf p}}
\newcommand{\bq}{{\bf q}}
\newcommand{\ag}{\delta_{\alpha\gamma}}
\newcommand{\bd}{\delta_{\beta\delta}}
\newcommand{\ad}{\delta_{\alpha\delta}}
\newcommand{\bg}{\delta_{\beta\gamma}}
\newcommand{\gc}{g_{c,2}}

\begin{document}
\title{Fermi liquid near Pomeranchuk quantum criticality}
\author{Dmitrii L. Maslov$^{1}$ and Andrey V. Chubukov$^{2}$}
\date{\today}
\begin{abstract}
We analyze  the 
behavior of an itinerant Fermi system 
near a charge nematic ($n=2$)  
Pomeranchuk instability 
 in terms of the Landau Fermi liquid (FL) theory.
 The main object of our study is  
the fully renormalized vertex function $\Gamma^\Omega$, related to the Landau
interaction function. We derive $\Gamma^\Omega$ for a model case of the long-range interaction
in the nematic channel.  Already within the Random Phase Approximation (RPA), the vertex  is singular near the instability.
The full vertex, obtained by resumming the ladder series composed of the RPA vertices, differs from the RPA result
 by a multiplicative renormalization factor $Z_\Gamma$, related to the single-particle residue $Z$ and effective mass renormalization $m^*/m$.
 We employ the Pitaevski-Landau
 identities, which express 
 the derivatives of the self-energy 
 in terms of $\Gamma^\Omega$, 
 to obtain and solve a
 set of coupled non-linear 
equations  for $Z_\Gamma$, $Z$, and $m^*/m$. We show that
near the transition the system enters 
a  critical FL regime, where
 $Z_\Gamma 
 \sim Z \propto (1 + g_{c,2})^{1/2}$  and $m^*/m \approx  1/Z$, where $g_{c,2}$ is the $n=2$ charge Landau component which approaches $-1$ at the instability. 
 We construct the Landau function of the critical FL and show that all but $g_{c,2}$  Landau components 
 diverge at the critical point. 
 We also show that in 
   the critical regime the
 one-loop 
 result for the self-energy 
 $\Sigma (K) \propto \int dP G(P) D (K-P)$ 
 is asymptotically exact  if
   one identifies
   the effective interaction 
   $D$
     with the RPA form of $\Gamma^\Omega$.
 \end{abstract}
\affiliation{
$^{1}$ Department of
Physics, University of Florida, P. O. Box 118440, Gainesville, FL
32611-8440
\\
$^{2}$ Department of Physics, University of Wisconsin-Madison, 1150 University
Ave., Madison, WI 53706-1390}
\maketitle
\section{Introduction}
The
 Fermi liquid (FL) theory  states that 
 low-energy excitations
 in a system of interacting  fermions
   are  represented by fermionic
quasiparticles which differ quantitatively but not
 qualitatively from
  free fermions. \cite{agd,pines,anderson,baym,shankar}
 Quantitative changes are 
   encoded 
   in the
Landau function, ${\hat g}$,
 which is a tensor in the spin space and a function of 
 the momenta of colliding quasiparticles in the orbital space. Angular harmonics of ${\hat g}$ in the charge and spin channels, $g_{c,n}$ and $g_{s,n}$,
 determine  renormalized values of
 various observables.

The FL theory also allows for
 instabilities 
  which occur
 either due to Cooper pairing
  or to  
 symmetry-breaking deformations of the Fermi surface (FS).
The latter are known as  Pomeranchuk instabilities.\cite{pomer}
In the FL notations,  
a Pomeranchuk instability
occurs when one of the Landau harmonics
 approaches $-1$.

 Examples of Pomeranchuk instabilities include phase separation $(g_{c,0} =-1$),  a ferromagnetic transition
   ($g_{s,0} =-1$),   at which 
   the FSs
  of spin-up and spin-down fermions split
  apart,    
 and nematic-type 
 transitions in the charge 
\cite{metzner_halboth,Oganesyan,yamase_1,MRA,yamase_2,Kee,andy,woelfle,senthil,neto,woelfle_2}
  and spin channels, \cite{hirsch,fradkin,spin_we} which lower the rotational symmetry of the FS.
In modern literature,
the point in the parameter space where a Pomeranchuk instability occurs is called a
 quantum critical point (QCP).
 
  As the system
 approaches a Pomeranchuk instability, a certain  \lq\lq FS susceptibility\rq\rq, which measures \lq\lq softness\rq\rq\ of the Fermi surface with respect to deformations of particular symmetry, diverges. \cite{pomer,MRA} In certain cases, e.g., for a ferromagnetic transition, a diverging FS susceptibility coincides with the thermodynamic susceptibility whose divergence signals a phase transition. In general, however, the FS and thermodynamic susceptibilities are different. For example, any instability resulting in a deformation of the FS in an isotropic FL is not accompanied by a divergence of a thermodynamic susceptibility, which remains isotropic in the disordered phase.  In those cases, a way to detect Pomeranchuk instabilities is to construct the correlation function of the incipient order parameter of certain symmetry.\cite{metzner_halboth}
 A susceptibility corresponding to this correlation function,
 $\chi_{a^*,n^*}$, diverges as $1/(1 + g_{a^*,n^*})$ at a QCP in the critical channel $\{a^*,n^*\}$, 
  where $a=c,s$.  Although it does not follow from any general 
 principle, it is usually assumed
  that 
Landau components in all other channels 
  are not affected by a Pomeranchuk instability in 
  the $\{a^*,n^*\}$ channel.

 Taken at
 face value, the
 last assumption implies that, near a Pomeranchuk QCP, 
the effective mass $m^* = m(1 + g_{c,1})$ remains finite. 
  However, 
  Hertz-Millis--type effective theories of Pomeranchuk QCPs,\cite{Lee,ccm,monien,Khvesh,aim,Oganesyan,Kee,MRA,cmgg,chub_cross,rech,cgy,CK06,Lawler,senthil,cm_chi,tigran,khodel}
 in which 
 the original four-fermion interaction
 is replaced  by the interaction of fermions with 
  fluctuations
 of the incipient order parameter,  predict  a different behavior.
 In these theories, the effective mass diverges at a QCP in dimensions $D \leq 3$:
 as $|\ln (1 + g_{a^*,n^*})|$ in $D=3$ and as
$(1 + g_{a^*,n^*})^{-1/2}$ in $D=2$.
 As the FL theory is supposed to be valid 
  at $T=0$
 down to 
   the very QCP, the divergence of $m^*$ implies 
    that of $g_{c,1}$,
 in clear disagreement with the assumption that 
 all but $g_{a^*,n^*}$ Landau components are not affected by the Pomeranchuk instability in a particular channel.  

   As  
 susceptibilities in 
 the
 FL theory contain the fully
 renormalized mass,
 it then becomes an issue whether 
  this singular mass renormalization should 
  be included
  into the ordinary FL formula for the susceptibility in the critical  $\{a^*,n^*\}$ channel $\chi_{a^*,n^*} = (1 + g_{c,1})/(1 + g_{a^*,n^*})$ and 
   also whether non-critical channels 
  can be \lq\lq dragged\rq\rq\/ to criticality simply due to a divergence in the effective mass. 

In this communication, we
re-visit  this issue.
We consider a
 Pomeranchuk QCP in the charge channel 
(a QCP in the spin channel deserves 
 a separate consideration because of subtle
 issues related to spin conservation, see
 Ref.~\onlinecite{spin_we}).
  For definiteness, we focus on the charge nematic instability in the $n=2$ channel (Refs.~\onlinecite{metzner_halboth,Oganesyan,yamase_1,MRA,yamase_2,Kee,andy,woelfle,senthil,neto,woelfle_2});
 our 
  conclusions,
  however,  are  also valid for all $n >1$. 
  We  show that, for $D \leq 3$
the assumption of only one critical Landau component
   holds only at some \lq\lq distance\rq\rq\/ from a QCP in the parameter space.
  In the immediate vicinity of a QCP,  
   a FL of new type  emerges. In what follows, we will refer to this new FL as to a \lq\lq critical FL\rq\rq\/. The interactions at high energies, which were the cause of the Pomeranchuk instability in the first place, play the role of \lq\lq bare\rq\rq\/ interactions for the critical FL. Accordingly, the enhanced nematic susceptibility $\chi^{\mathrm{FL}}_{c,2} \propto 1/(1 + g_{c,2})$ is a bare susceptibility of the critical FL. The low-energy interactions, mediated by soft collective fluctuations  in the $n=2$ charge channel, lead to further renormalizations of the FL parameters. These renormalizatons are encapsulated in the \lq\lq critical \rq\rq\/Landau function, ${\bar g}$, which we show to have
both charge and spin components  with \emph {any} $n$, even if the original, \lq\lq high-energy\rq\rq\/ FL has only the $n=2$ charge Landau component.

Our key result is that
 {\it all} components of ${\bar g}$ diverge
 in the same way, i.e.,  as $1/(1 + g_{c,2})^{1/2}$, upon approaching a  nematic QCP in $D \leq 3$.    In particular,
  a divergence of ${\bar g}_{c,1}$ implies that 
$m^*/m = 1 + {\bar g}_{c,1}$ diverges at the
 QCP as well. At the same time, 
since divergences are the same for all ${\bar g}_{a,n}$, 
they cancel out in the expressions for the susceptibilities 
$\chi_{a,n} \propto (1 + {\bar g}_{c,1})/(1 + {\bar g}_{a,n})$, which
  retain the same values $\chi^{\mathrm{FL}}_{a,n}$ as in the  original  FL. 
In particular, 
 susceptibilities in channels different from the critical one 
 remain finite at a QCP, while $\chi_{c,2}$ 
preserves its $1/(1+g_{c,2})$ form. This means
 that  the divergence of the effective mass near a Pomeranchuk QCP
 does {\it not} affect  the behavior of  any of the susceptibilities.

To obtain these results, we 
derive diagrammatically an
expression for the 
fully renormalized anti-symmetrized interaction vertex 
${\hat\Gamma}^\Omega (\mathbf{k},\mathbf{p})$  between the particles 
with momenta
$\mathbf{k}$ and $\mathbf{p}$ 
on
 the Fermi surface. This  vertex  is 
 obtained from a more general vertex function in the limit of zero momentum transfer and vanishing frequency transfer and 
  is the ``input'' parameter for the 
 FL theory:
  the Landau function is proportional to ${\hat\Gamma}^\Omega$
\beq
{\hat g} = 2 \nu Z^2 \frac{m^*}{m} {\hat\Gamma}^\Omega,
\label{y_1}
\eeq
 where $Z$ is the quasiparticle residue. 

We identify the most relevant part of 
 ${\hat\Gamma}^\Omega$
 near a QCP and show that it
describes 
an interaction mediated by soft 
collective bosonic fluctuations in the $n=2$ charge channel.
For  small $|\mathbf{k}-\mathbf{p}|
 \approx  \kf  \theta$,  we find 
 $\Gamma^\Omega (\theta) \propto 1/\left[ 1 + g_{c,2} + (a\kf  \theta)^2 + ..\right]$ where dots stand for less relevant terms. The length scale $a$ is the effective radius of the interaction $U(q)$ in the $d-$wave charge channel, and the product $a\kf$ is a dimensionless parameter of our theory. The calculations are under control if $a\kf \gg1$, which we assume to hold. For $a\kf \gg 1$, 
  the system is the 
 critical FL regime 
 when $(1 + g_{c,2}) < 1/(a\kf)^2\ll 1$.  

Such a 
 singular form of 
${\hat\Gamma}^\Omega$
was proposed earlier on phenomenological grounds\cite{pheno} and obtained within the Random Phase Approximation (RPA) for the Hubbard model 
near an antiferromagnetic instability.\cite{scal} To obtain the full ${\hat\Gamma}^\Omega$ in our case, we first generalize the RPA result to the $n=2$ charge
 Pomeranchuk instability. We show that the RPA-type formula 
 for the effective interaction 
 ${\hat\Gamma}^{\mathrm{RPA}}(\bk,\bp) \propto 1/(1 - U ({\bf k} - {\bf p}) 
\Pi_2 (\bk,\bp))$, where  $\Pi_d (\bk,\bp)$ is the 
static polarization bubble in the $n=2$ channel, 
is reproduced by summing up
 the  diagrams which do not renormalize the bare interaction $U$. 
 
Next, we analyze the diagrams for ${\hat\Gamma}^\Omega$ beyond RPA and 
 show that in the critical FL regime
the full $ \Gamma^\Omega$ differs from 
 $\Gamma^{\Omega, \mathrm{RPA}}$ 
 by a 
 constant factor $Z^{-1}_\Gamma \sim  Z^{-1}$.
  This relation
 results 
 from re-summation of a particular non-RPA series of diagrams
 which includes renormalization of $U$ into a full dynamic interaction.
 We show that all other non-RPA diagrams are relatively small in powers 
of $1/a\kf$ and can therefore be neglected.  The existence of an extra factor $Z_\Gamma$ between  
 ${\hat \Gamma}^\Omega$ and ${\hat\Gamma}^{\Omega, \mathrm{RPA}}$ is very
 important  for our analysis -- 
just the RPA form of ${\hat\Gamma}^\Omega$ would produce nonsensical results in the FL description.
  
Another input parameter for a FL theory is the quasiparticle residue $Z$.
To obtain $Z$, we use the exact Pitaevski-Landau relations \cite{agd,lp} which express $Z$ in 
 terms of ${\hat\Gamma}^\Omega (\bk,\omega_k;\bp,\omega_p)$, where now $\bp$ and $\bk$ are not necessarily at the FS and 
 $\omega_{k,p}$ are finite. We extend the previous calculation of ${\hat\Gamma}^\Omega$ to fermions away from the FS and obtain $Z$ as a function of 
$(1 + g_{c,2})^{1/2}$ with $a\kf$ as a parameter. This function is rather 
complex but reduces to  a simple form $Z \sim  (a\kf) (1 + g_{c,2})^{1/2}$ in the critical FL regime. Using this form of $Z$ and full
${\hat\Gamma}^\Omega$,
 we  construct the Landau function of the critical FL and 
show that all its components diverge in a way discussed above.
We also
show that the effective mass $m^*$ diverges at but not before a QCP.
 In this respect, our results 
do not support the conjecture\cite{khodel}
 that the effective mass may diverge before the system reaches a QCP. 

We also discuss the relation between the exact formula for the self-energy $\Sigma (k, \omega)$ to 
linear order in $\omega$ and $k-\kf$,  obtained 
from the Pitaevski-Landau relations, and a one-loop formula for  $\Sigma$ due to an exchange by soft 
bosonic collective excitations. We show that the one-loop formula is 
 asymptotically exact in the critical FL regime if  the effective interaction is replaced 
 by $Z{\hat\Gamma}^\Omega$, i.e.,  by the RPA form of the effective vertex  ${\hat\Gamma}^{\Omega,\mathrm{RPA}}$. 
 The corrections to the 
 one-loop formula are small in 
 $1/a\kf$ and in $|\omega|/
  \omega_{\mathrm {FL}}
  $, where $\omega_{\mathrm{FL}} \propto (1 + g_{c,2})^{3/2}$ is the 
 upper
  boundary of the FL behavior.
We emphasize that the one-loop approximation is valid only for the linear-in-$\omega$ term in the self-energy. The next,
$\omega^2 \ln\omega$ term has contributions from all orders even 
if the  theory is extended 
 to a large number of fermionic flavors $N$. \cite{sslee_1} 
 Outside 
 the FL regime, i.e,  for
 $|\omega| > \omega_{\mathrm{FL}}$, the self-energy scales as $\omega^{2/3}$.
 Some vertex corrections in this regime are small in $1/N$
 (Refs. \onlinecite{aim,rech,cm_chi,tigran}), while others
 remain $O(1)$ even  for large $N$ (Ref. ~\onlinecite{sslee}).

Regarding the full form of the self-energy near a QCP, we show that,
for  $1 + g_{c,2} \ll 1$, the self-energy is \lq\lq local\rq\rq\/, in the sense that it 
 depends primarily on $\omega$ but 
 not on $\epsilon_k=\vf(k-\kf)$.
 The prefactor for the $\epsilon_k$ term in the self-energy
 scales as $1/a\kf$  and is, therefore, small.

A Pomeranchuk instability in the $d-$wave charge channel was 
 introduced in the context of the RG analysis of potential 
 instabilities of a 2D Hubbard model.
\cite{metzner_halboth,yamase_1} Shortly thereafter, Oganesyan et al. analyzed in detail a $d-$wave charge instability in isotropic systems.\cite{Oganesyan} 
The subject
 attracted substantial interest\cite{MRA,yamase_2,Kee,andy,woelfle,senthil,neto,woelfle_2} 
 both from the theoretical perspective and also due to
 potential relevance to cuprates\cite{hinkov}
and ruthenites.\cite{sr}  
There are some subtle 
 differences between a $d-$wave Pomeranchuk instability in lattice systems\cite{metzner_halboth,yamase_1,MRA,yamase_2} and 
an $n=2$ Pomeranchuk instability in isotropic systems,\cite{Oganesyan,Kee,andy,woelfle,senthil,neto} 
but our conclusion is 
that the physics does not change qualitatively 
between isotropic and lattice cases (see below).      

Properties of the Fermi liquid near a charge Pomeranchuk instability were 
studied by Rosch and W{\"o}lfle.
 \cite{woelfle} They did not consider the Landau function near a nematic QCP, 
but obtained the effective mass in terms of  the critical parameter
$1 + g_{c,2}$ using the one-loop approximation  for the self-energy
with an RPA form of the effective vertex. 
Our results agree with  Ref. ~\onlinecite{woelfle} in that the effective mass can be obtained within the one-loop approximation with the RPA vertex, but our dependence of the mass  on  $1 + g_{c,2}$ is different from that in Ref.~\onlinecite{woelfle}. The disagreement originates from the difference in the forms of the static polarization bubble for dressed fermions, which we discuss 
in Sec.~\ref{sec:away}.

The structure of the paper is as follows. 
 In Sec.~\ref{sec:2}, we review a general procedure of constructing the FL vertices.
Section ~\ref{sec:CFLT} is devoted to the critical FL theory. In Sec. ~\ref{sec:nematic}, we
introduce a model for the nematic Pomeranchuk instability in the charge channel. In Sec. ~\ref{sec:nematic_RPA},
we obtain the FL vertex  ${\hat\Gamma}^\Omega$ near this 
 instability 
within 
 the 
RPA approximation. In Sec. \ref{sec:3_1}, we go beyond the RPA level and 
obtain 
 the full  vertex ${\hat\Gamma}^{\Omega}$.
 In Sec. \ref{sec:4},
we obtain the quasiparticle residue $Z$, the vertex renormalization $Z_{\Gamma}$, and effective mass $m^*$  from the exact Pitaevski-Landau relations.
 In the same Section, we also analyze 
the crossover between ordinary and critical FLs.
 In Sec.\ref{sec:3}, 
  we obtain the Landau function in the critical FL
 using the full vertex found in Sec.~\ref{sec:3_1} and show that all components of this Landau function diverge at a QCP. This is the main result of our paper.
In Sec.\ref{sec:4a}, we discuss the relation between the one-loop and exact Pitaveskii-Landau forms of the self-energy near the FS.
We present our concluding remarks in Sec.  \ref{sec:5}.    
A number of technical issues are discussed in Appendices \ref{app:b}, \ref{app:a}, and \ref{sec:subtle}.

\section{Diagrammatic description of an ordinary Fermi liquid}
\label{sec:2}
We begin with a brief overview of the diagrammatic description of an ordinary FL.  
The FL theory describes effects of the interactions between fermions confined to a near vicinity of the FS. The interactions which involve 
 fermions away from the Fermi surface are absorbed into 
the Landau function $g_{\alpha\beta,\gamma\delta}$. 
This function is related to an exact antisymmetrized vertex $\Gamma^\Omega_{\alpha\beta,\gamma\delta}$
 via $g_{\alpha\beta,\gamma\delta} = 2 \nu Z^2 (m^*/m) \Gamma^\Omega_{\alpha\beta,\gamma\delta}$ ($\nu=m/2\pi$ is the density of states in 2D). 
The vertex $\Gamma^\Omega_{\alpha\beta,\gamma\delta}$
is defined  for particles at the FS in the limit of
zero momentum transfer $q$ and vanishing 
 energy transfer $\Omega$ (Ref.~\onlinecite{agd})
\begin{widetext}
\beq\Gamma^\Omega_{\alpha \beta; \gamma \delta}  =\lim_{q/\Omega\to 0}\left[
\Gamma({\bf k},0;{\bf p},0\vert {\bf k}-{\bf q},-\Omega; {\bf p}+{\bf q},\Omega) \delta_{\alpha \gamma} \delta_{\beta \delta}  -
\Gamma({\bf k},0;{\bf p},0\vert {\bf p}+{\bf q},\Omega;{\bf k}-{\bf q},-\Omega) \delta_{\alpha \delta} \delta_{\beta \gamma}\right]\vert_{k=p=\kf },\label{gammaomega}
\eeq
\end{widetext}
where $\Gamma({\bf k},\omega_k;{\bf p},\omega_p\vert{\bar {\bf k}},{\bar\omega}_k;{\bar{\bf p}},{\bar \omega}_p)$ is an exact nonsymmetrized vertex. 
 The momentum transfer is equal to ${\bf q}$ in the first term of Eq.~(\ref{gammaomega}) and to ${\bf k}-{\bf p}-{\bf q}$ in the second one. 

 For a  generic ratio of $q$ and    $\Omega$, 
diagrams for the vertex 
$\Gamma({\bf k},\omega_k;{\bf p},\omega_p\vert{\bar {\bf k}},{\bar\omega}_k;{\bar{\bf p}},{\bar \omega}_p)$ [and its exchange counterpart  $\Gamma({\bf k},\omega_k;{\bf p},\omega_p\vert{\bar{\bf p}},{\bar \omega}_p;{\bar {\bf k}},{\bar\omega}_k)$] can be separated into two groups depending on whether
  they contain propagators of \lq\lq soft\rq\rq\ particle-hole pairs
\beq
 {\cal P} \left(\mathbf{q},
 \Omega,  \mathbf{\hat{k}}\right)= -\nu\int d\epsilon_{k}\int  \frac{d\omega}{2\pi} G\left(\mathbf{k}+\mathbf{q},\omega+
 \Omega
 \right)G\left(\mathbf{k},\omega\right)
 \label{calp}
 \eeq
 with  $q \to 0$ and $\Omega \to 0$, where $\epsilon_k=\vf(k-\kf)$. 
  The meaning of ${\cal P}$ is the propagator of a particle-hole excitation
  with momentum $\mathbf{q}$ and energy $\Omega$ formed by fermions moving in the direction of $ \mathbf{\hat{k}}=\mathbf{k}/k$.
 
 For vanishingly small $q$ and $\Omega$, 
 the dynamic part of the bubble ${\cal P}$ is determined by fermions on the FS;
therefore, diagrams with soft bubbles are 
  attributes of the FL theory. For these diagrams, the order of limits $q\to 0$ and $\Omega\to 0$ matters because
   ${\cal P}=0$ in the limit of $q/\Omega\to 0$ and
is finite (equal to $\nu$) in the limit 
$\Omega/q\to 0$. The diagrams without soft bubbles generally involve 
 fermions with high energies, of order $\ef$. For these diagrams, the order of limits $q\to 0$ and $\Omega\to 0$ is irrelevant. 

\begin{figure}[tbp]
\centering
\includegraphics[width=1.0\columnwidth]{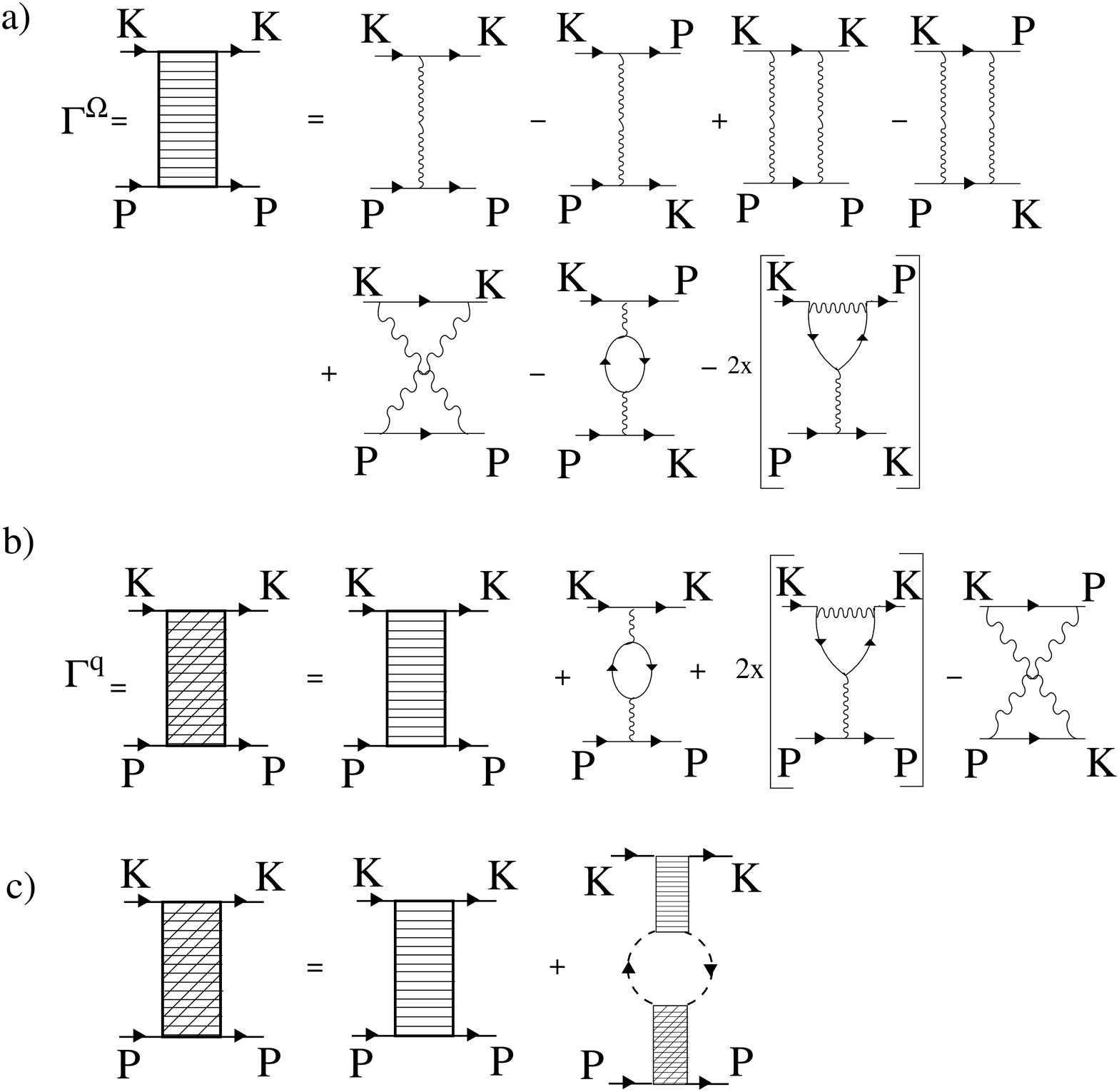}
\caption{Fermi-liquid vertices to second order in the bare interaction (wavy line). \emph{a}) 
 Diagrams for $\Gos$. \emph{b}) Diagrams for $\Gamma^q_{\alpha\beta;\gamma\delta}$. \emph {c}) Diagrammatic representation of the relation between $\Gos$  and $\Gqs$, 
Eq.~(\ref{qomega}). A bubble composed of dotted lines represents the particle-hole propagator ${\cal P}$ [see Eq.~(\ref{calp})] in the limit $\Omega/q\to 0$, where ${\cal P}$ reduces to a constant.
}  
\label{fig:gammaomega}
\end{figure}

Because $\Gamma^\Omega_{\alpha\beta;\gamma\delta}$ is the vertex in the limit 
$q/\Omega \to 0$, it  does not include 
soft particle-hole bubbles (${\cal P} =0$ in this limit)
 and, in that sense, it is a high-energy property
 playing a role of the  bare vertex 
in the FL theory.  The second-order diagram for 
 $\Gamma^\Omega_{\alpha\beta;\gamma\delta}$ are shown in Fig.~ \ref{fig:gammaomega}\emph{a}). The wavy line 
 in these diagrams is the bare, static
 interaction potential $
  V(\bk,\bp;{{\bar \bk}},{{\bar \bp}})
 $   which may depend not only on the momentum transfer 
$\bk-{\bar\bk}$ but also on the incoming momenta 
 themselves.

On the other hand, physical observables, e.g., the specific heat or the 
 nematic charge susceptibility,
  contain  contributions from soft bubbles  because the  observables are affected by  elastic
collisions between particles 
right on the FS, i.e., by processes 
with
 $\Omega =0$ and finite $q$.
  The corresponding antisymmetrized vertex is called $\Gamma^q_{\alpha\beta;\gamma\delta}$.  In Fig. \ref{fig:gammaomega}\emph{b}),
   we show  the second-order diagrams for  $\Gamma^q_{\alpha\beta;\gamma\delta}$. 
The full  vertex $\Gamma^q_{\alpha\beta;\gamma\delta}$ is  obtained from $\Gamma^\Omega_{\alpha\beta;\gamma\delta}$ by summing up an infinite  series of ladder 
diagrams shown in Fig. \ref{fig:gammaomega}\emph{c}), 
  which leads to a familiar integral equation\cite{agd} 
\bwt
\beq
\Gqs(K_F,P_F)=\Gos(K_F,P_F)-Z^2\nu\int \frac{d\theta}{2\pi}\Gamma^{\Omega}_{\alpha\xi;\gamma\eta}(K_F,K^\prime_F)\Gamma^q_{\eta\beta;\xi\delta}(K^\prime_F,P_F),
\label{qomega}
\eeq
\ewt
where a shorthand $K_F, P_F$ denotes the \lq\lq four-momentum\rq\rq\/ on the FS, i.e.,  
$K_F\equiv\{\bk_{\mathrm{F}},\omega_k=0\}$, 
 $\bk_{\mathrm{F}}\equiv \kf{\hat k}$, and $\theta$ is the angle between ${\hat\bk}$ and ${\hat\bk^\prime}$.

  The full vertex $\Gamma^q_{\alpha\beta;\gamma\delta}$  is related to the scattering 
 amplitude  $f_{\alpha\beta;\gamma\delta}$ in the same way as 
$\Gamma^\Omega_{\alpha\beta;\gamma\delta}$  is related to $g_{\alpha\beta;\gamma\delta}$, i.e., via $f_{\alpha\beta;\gamma\delta} = 2\nu Z^2 (m^*/m) 
\Gamma^q_{\alpha\beta;\gamma\delta}$. The relation between ${\hat f}$ and ${\hat g}$ 
 takes a particularly simple form when expressed via partial components in the charge and spin channels,
$f_l$ and $g_l$:
 \beq
f_{a,l} = \frac{g_{a,l}}{1 + g_{a,l}}.
\label{y_2}
\eeq
 Using these relations, one  obtains
 FL formulas for observables in terms of $g_{a,l}$. 

\section{Critical Fermi-liquid  theory}
\label{sec:CFLT}
\subsection{Nematic charge fluctuations}
\label{sec:nematic}
We follow early work\cite{Oganesyan,MRA,Kee,neto,woelfle,woelfle_2} and consider   
 a nematic charge instability described by the model Hamiltonian with 
the interaction in the $d-$wave charge channel
\beq
H=\sum_{\bk,\bp,\bq}U(\bq)d_{\bk}d_{\bp}c^{\dagger}_{\bk+\bq/2,\alpha}c^{\dagger}_{\bp-\bq/2,\beta}c_{\bp+\bq/2,\beta}c_{\bk-\bq/2,\alpha}.
\label{Hdwave}
\eeq
Here \beq
d_{\bk}=\sqrt{2}\cos(2\phi_{\bk})
\label{df}
\eeq is the \lq\lq$d$-wave\rq\rq formfactor and $\phi_{\bk}$ is the angle between $\bk$ and an arbitrarily chosen $x$-axis. Hamiltonian (\ref{Hdwave}) describes the interaction between nematic fluctuations of the electron density 
$\rho_d(\bq)=\sum_{\bk}d_{\bk}c^{\dagger}_{\bk+\bq/2,\alpha}c_{\bk-\bq/2,\alpha}$.

To keep the treatment under control, we assume that the interaction $U(\bq)$ is sufficiently long-ranged in real space, i.e., 
\beq
U(\bq)=U_0 P(qa),
\eeq
where the function $P(x)$,  subject to  $P(0)=1$, is a decreasing function of its argument, 
 and the effective interaction radius $a$ is much larger than the inverse average distance between particles, i.e. $\kf a\gg 1$. 
We also assume that $U(\bq)$ is analytic for small $q$:
\beq
U(\bq)=U_0(1-(qa)^2+\dots).
\label{int}
\eeq

Equation (\ref{Hdwave}) is a reduced version of a more general
 interaction between  quadrupolar fluctuations of the electron density
$c^{\dagger}_{i,\alpha} Q_{ij}c_{j,\alpha}$, where $Q_{ij} = \delta_{ij} \nabla^2 - 2 \partial_i \partial_j$, and $i,j =x,y$ (Refs.~\onlinecite{Oganesyan,woelfle_2}). Such an interaction can be decoupled into a longitudinal part, which is the same as in Eq.~(\ref{Hdwave}), and a transverse part with the momentum-dependent factor ${\bar d}_{\bk}{\bar d}_{\bp}$, where ${\bar d}_{\bk}=\sqrt{2}\sin(2\phi_{\bk})$. The two terms contribute separately to the fermionic self-energy and $\Gamma^\Omega$ and can be treated independently of each other. 
To simplify the presentation, 
 we consider only the longitudinal part of the quadrupole interaction, given by Eq.~(\ref{Hdwave}).
The effect of the transverse part on thermodynamic properties of a FL has recently been considered by Zacharias, Garst, and W{\"o}lfle
\cite{woelfle_2}.
 
The nematic charge susceptibility 
of free fermions is defined as
\bea
&&\chi^{(0)}_{c,2}(\bq,\Omega)=2\Pi_d(\bq,\Omega)\\
&&=-2\sum_{\bk,\omega_k}d_{\bk}^2 G(\bk+\bq/2,\omega_k+\Omega/2)G(\bk-\bq/2,\omega_k-\Omega/2).\nn
\eea
We will be interested in the long wavelength and low frequency $d$-density fluctuations with $q\ll \kf$ and $|\Omega|/\vf q\ll 1$. In this regime, $\chi^{(0)}_{c,2}$ for free 2D 
 fermions with quadratic spectrum $\varepsilon_{\bk}=k^2/2m$ is
\bea
\chi^{(0)}_{c,2}(\bq,\Omega)
 =2\nu\left(1-\frac{q^2}{2\kf ^2}-
  2\cos^2(2\phi_{\bq})\frac{|\Omega|}{\vf q}+\dots\right).
\label{pidfree}
\eea
Notice that Landau damping of nematic fluctuations is anisotropic whereas the dispersion of static $\chi^{(0)}_{c,2}$ with $q$ (absent for $n=0$ density fluctuations for $q\leq 2\kf$) is  isotropic.

To first order in the interaction, 
 $\Gos$ is given by the first two diagrams in Fig.~\ref{fig:gammaomega} {\it a}):
\beq
\Gamma^\Omega_{\alpha\beta;\gamma\delta}(\bk,\bp) =U_0 d_{\bk}d_{\bp} \ag\bd-U(\bk-\bp)d_{\frac{\bk+\bp}{2}}^2\ad\bg .
\label{t_6_1}
\eeq
According to our definition of $d_{\bk}$, when both $\bk$ and $\bp$ are on the FS,
\bea
d_{\frac{\bk+\bp}{2}}^2&=&\cos(2\phi_{\bk})\cos(2\phi_{\bp})+1-\sin(2\phi_{\bk})\sin(2\phi_{\bf p})\notag\\
&=&\frac{1}{2}d_{\bk}d_{\bp}+\dots\label{dk}
\eea
where dots stand for non-$\cos( 2 \phi_{\bk})$ terms which we neglect.
 Using Eq.~(\ref{dk}) and the $SU(2)$ identity
\beq
\ad\bg=(\ag\bd+{\vec\sigma}_{\alpha\gamma}\cdot{\vec\sigma}_{\beta\delta})/2
\label{su2},
\eeq
 we separate Eq.~(\ref{t_6_1}) into the charge and spin channels as
\bea
&&\Gamma^\Omega_{\alpha\beta;\gamma\delta}(\bk,\bp) =\Gamma_c^{\Omega}\delta_{\alpha \gamma} \delta_{\beta\delta}+\Gamma^{\Omega}_s{\vec \sigma}_{\alpha\gamma}\cdot {\vec\sigma}_{\beta \delta}\label{t_6a}\\
&&\Gamma^{\Omega}_c=d_{\bk}d_{\bp}\left[U_0-\frac{1}{4}U(\bk-\bp)\right];\;
\Gamma^{\Omega}_s= - \frac{1}{4}d_{\bk}d_{\bp}U(\bk-\bp).\nn
\eea
At this stage, the interaction is static and mostly in the d-wave channel. 
Hence the effective mass equals to the bare one and the quasiparticle residue $Z=1$. The Landau function is then obtained by simply 
 multiplying Eq.~(\ref{t_6a}) by  $2\nu$. 
 Since, by assumption, $U(\bk-\bp)$ is peaked at $\bk=\bp$, the charge $d-$wave Landau harmonic, $g_{c,2}$,  is much larger than the spin 
 $d$-wave
 harmonic $g_{s,2}$.
  Indeed, for $a\kf \gg 1$,
\bea
g_{c,2}&=&2\nu\int\int\frac{d\phi_{\bk}}{2\pi}\frac{d\phi_{\bp}}{2\pi}d_{\bk}d_{\bp}\Gamma^{\Omega}_c
\approx 2\nu U_0\left(1-\frac{c}{a\kf }\right)\notag\\
g_{s,2}&=&2\nu\int\int\frac{d\phi_{\bk}}{2\pi}\frac{d\phi_{\bp}}{2\pi}d_{\bk}d_{\bp}\Gamma^{\Omega}_s
\approx -2\nu U_0\frac{c}{a\kf } \ll g_{c,2},\nn\\
 \label{first}
\eea
where $c=(3/4\pi)\int^{\infty}_0dxP(x)$.
 To leading order in $1/\kf a$, we then have $g_{c,2} = 2 \nu U_0$, $g_{s,2} =0$.
We see that $g_{c,2}$ approaches $-1$  when
 $U_0$  approaches $ -1/(2\nu)$.

With only $g_{c,2}$ being non-zero, the
 nematic spin susceptibility $\chi_{s,2}=\chi^{(0)}_{s,2}(1+g_{c,1})/ (1+g_{s,2}) = \chi^{(0)}_{s,2}$  retains its bare value,  while the charge susceptibility 
 is
\beq
\chi_{c,2}=\chi^{(0)}_{c,2} \frac{1+g_{c,1}}{1+g_{c,2}} = \frac{\chi^{(0)}_{c,2}}{1 + 
 g_{c,2}}
\label{y_6}
\eeq
 diverges when $g_{c,2}$ tends to $-1$. 
This corresponds to the ``conventional'' scenario of a Pomeranchuk QCP, where only one of the Landau components approaches $-1$ but the effective
mass does not diverge. We will see below that such a behavior does not survive in a non-perturbative theory with fully renormalized vertices.
\begin{figure}[tbp]
\centering
\includegraphics[width=1.0\columnwidth]{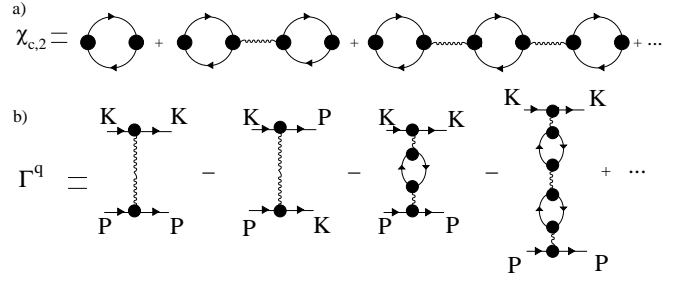}
\caption{RPA series for the nematic susceptibility [\emph{a})] and 
 $\Gqs$ [\emph{b})].  Solid circles represent  $d$-wave vertices 
 $d_{\bk}$ and wavy lines represent the forward-scattering part of the interaction, $U_0$.
 The bubbles are with zero vanishing external momentum and zero frequency --the same bubbles as in Fig.~\protect\ref{fig:gammaomega} {\it b},{\it c}. 
Since, by assumption 
$a\kf\ll 1$, only the diagrams that contain $U_0$ are included.}  
\label{fig:dwave_1}
\end{figure}

Before 
 moving to the analysis of vertex renormalization,
  we note that 
Eq.~(\ref{y_6}) and the FL relation (\ref{y_2}) can be reproduced diagrammatically
by calculating directly the charge susceptibility
 by calculating  $\Gqs$.  In both cases we need to 
include  only those diagrams that contain $\Pi_d ({\bf q}, \Omega)$ at
 vanishing momentum and 
  zero
 frequency and, thus,
  do not contribute to renormalization of  $\Gos$. 
Summing up the ladder series
for $\chi_{c,2}$ in Fig.~\ref{fig:dwave_1} \emph{a}, we obtain  
\beq
\chi_{c,2}(\bq\to 0,\Omega=0) = \frac{\chi^{(0)}_{c,2}}{1 + 2 \nu U_0}= \frac{\chi^{(0)}_{c,2}}{1 + g_{c,2}},
\label{chi_RPA}
\eeq
 which coincides with Eq.~(\ref{y_6}). 
 The ladder series for  $\Gqs$ 
 in  Fig.~\ref{fig:dwave_1} \emph{b} yields
\beq
\Gamma^q_{\alpha\beta;\gamma\delta} = d_\bk d_\bp \left[\frac{U_0}{1 + 
 2 \nu U_0}\right] \delta_{\alpha\gamma} \delta_{\beta \delta}. 
 \label{t_5}
\eeq 
 Using the relation between 
 $\Gqs$ and the scattering amplitude $f_{\alpha \beta, \gamma \delta} = 2\nu \Gamma^q_{\alpha \beta, \gamma \delta}$, we 
 find
\beq
f_{c,2} \approx \frac{2 U_0 \nu}{ 1 +2 U_0 \nu}.
\label{y_7}
\eeq
Comparing Eqs.~(\ref{first}) and (\ref{y_7}), we see that 
 the FL relation
\beq
f_{c,2} = \frac{g_{c,2}}{1 + g_{c,2}}.
\label{t_8}
\eeq    
is indeed reproduced.
\subsection{Random Phase Approximation for nematic instability}
\label{sec:nematic_RPA}

\begin{figure}[tbp]
\centering
\includegraphics[width=1.0\columnwidth]{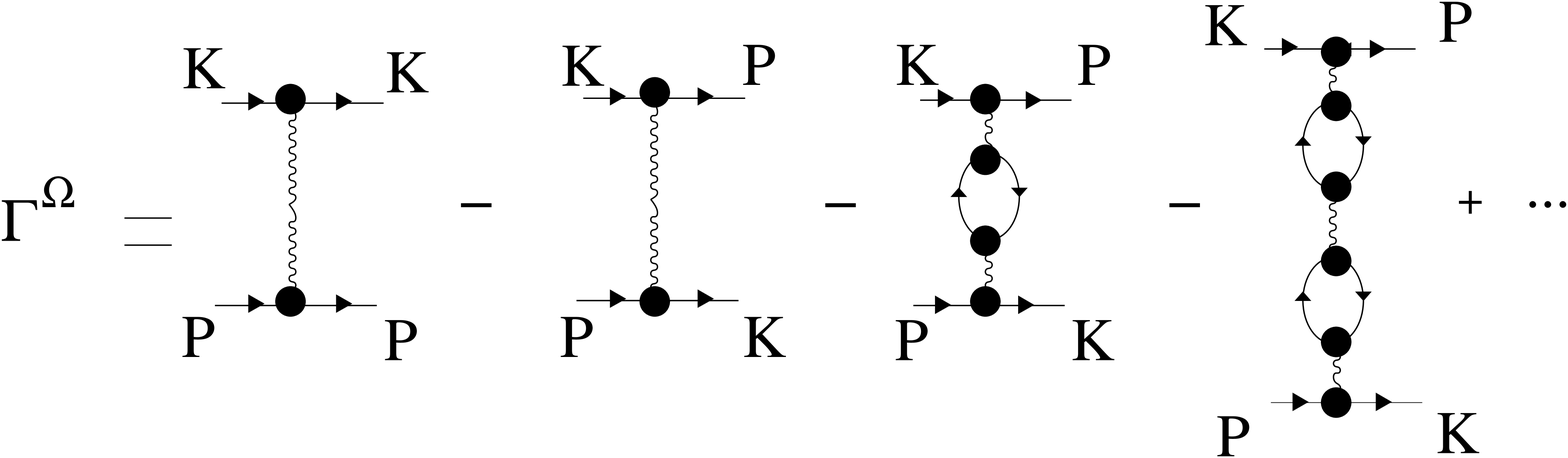}
\caption{RPA diagrams for $\Gos$. 
 The difference with the series for $\Gqs$ in Fig. \protect\ref{fig:dwave_1}{\it b} 
 is in that the polarization bubbles are evaluated at  finite external momentum ${\bf k} - {\bf p}$.}  
\label{fig:dwave}
\end{figure}

At the next step, we consider renormalization of $\Gamma^\Omega_{\alpha\beta;\gamma\delta}(\bk, \bp)$. An assumption of the long-range interaction in the nematic channel allows one to use the RPA, i.e., to retain only diagrams 
 with a maximum number of polarization bubbles $\Pi_d ({\bf q}, 0)$ at any given order, now with finite $\bq = \bk -\bp$.
 While all such diagrams contain the interaction only in the form 
 $U(\bk -\bp)$, non-RPA diagrams involve integrals of $U$ over 
intermediate momenta. Since each of these integrals contributes a small factor of $1/a\kf$ [see Eq.~(\ref{first})], non-RPA
 diagrams can be safely neglected. Summing up the RPA series  
 and neglecting the first order $U_0$ term, which is irrelevant near a QCP, we obtain
 \beq
\GRPA(\bk,\bp)\approx - \frac{d_{\bk}d_{\bp}}{2}~
\frac{U(\bk-\bp)}{1+2U(\bk-\bp)\Pi_d(\bk-\bp,0)}
\delta_{\alpha \delta} \delta_{\beta\gamma}. 
\label{grpa}
\eeq
Notice a ``wrong'' combination of spin indices ($\delta_{\alpha \delta} \delta_{\beta\gamma}$ instead of $\delta_{\alpha \gamma} \delta_{\beta\delta}$). 

Thermodynamic parameters of a FL (including the effective mass) are determined by $\GRPA(\bk,\bp)$ for the particles on the FS and 
at zero frequencies. However, for the consideration in the next sections, we will also need to know $\GRPA$ away from the FS, i.e., for momenta different from $\kf$ and for non-zero frequencies of incoming 
 and outgoing
fermions. A simplification is that,  for $1 + g_{c,2} \ll 1$, 
 relevant momenta are still close to $\kf$, while the relevant values of 
$|{\bf k} - {\bf p}|$ and  $\omega_k$, $\omega_p$ are small.  In this situation,
 $\GRPA$ is still given by an expression similar to Eq.~(\ref{grpa}) 
\beq
\GRPA(\bk,\omega_k; \bp,\omega_p)= 
 - \frac{1}{2}
 \frac{d_{\bk}d_{\bp} \ad\bg U(\bk-\bp)}{1 +  2 U(\bk-\bp)
 \Pi_d (\bk-\bp ,\omega_k-\omega_p)} 
\label{wed_5}
\eeq
but where
 now  $\Pi_d (\bk-\bp ,\omega_k-\omega_p)$ is  a dynamic bubble.

\subsubsection{The polarization bubble}
\label{sec:away}

The $d$-wave bubble of free fermions 
for $q\ll \kf$ and $|\Omega|\ll \vf q$:
 is given by Eq.~(\ref{pidfree}).
 Substituting this form into Eq.~(\ref{wed_5}),  replacing 
$U({\bf k} - {\bf p})$ by $U_0$ in the numerator, and neglecting the $q^2/2 k^2_F$ term compared to the $(a q)^2$ one which comes from 
 the expansion of 
$U({\bf k} - {\bf p})$ in Eq.~(\ref{int}), we obtain
\bwt
\bea
\GRPA(\bk,\omega_k; \bp, \omega_p) &=&d_{\bk}d_{\bp}\frac{1}{4\nu}\frac{1}{1+g_{c,2}+|\bk-\bp|^2a^2 + 
 2\cos^2 (2 \phi_{\bq}) \frac{|\omega_k - \omega_p|}{\vf |\bk-\bp|}}\ad\bg \nonumber \\
&& =d_{\bk}d_{\bp}
\frac{1}{8\nu}\frac{\ag\bd+{\vec\sigma}_{\alpha\gamma}\cdot{\vec\sigma}_{\beta\delta}}{1+g_{c,2}+|\bk-\bp|^2a^2 + 
 2\cos^2 (2 \phi_{\bq})\frac{|\omega_k - \omega_p|}{\vf |\bk-\bp|}},
\label{grpaqcp}
\eea
\ewt
where, as before, $\phi_{\bq}$ is the angle between ${\bf q}$ and an arbitrary chosen x-axis. 

This form of $\GRPA(\bk,\bp)$ is only valid, however,  at some distance away from a QCP, where fermions behave as nearly free quasiparticles. Near a QCP, the quasiparticle mass and residue 
 differ from the free-fermion values and $\Pi_d (\bq, \Omega)$ is to be treated as a fully renormalized bubble, 
which we label as  $\Pi_d^*(\bq,\Omega)$. 

Renormalization of the polarization bubble is a subtle issue and we pause here to discuss it in some detail.
  In general, the polarization bubble contains two types of renormalizations: the self-energy insertions, which transform
  the bare Green's functions into  the renormalized ones, and vertex corrections.  We will show in Sec.~\ref{sec:vertex} that the vertex corrections are small in the quasistatic limit ($\Omega\ll \vf q$); thus, in this limit, the bubble can be computed as a convolution of two renormalized Green's functions.  Still, one has to be careful even with this computation because the quasiparticle 
residue $Z$ and the effective mass $m^*$ depend on the
 energy they are measured at: near the FS, they approach the renormalized values of $Z<1$ and $m^*>m$, while at higher energies they approach the free values
  $Z =1$ and $m^*=m$. 
It turns out that renormalizations of the three terms in the expansion of the free bubble [Eq.~(\ref{pidfree})] are determined by different energies. First, we consider the constant term $\Pi_d^*(\bq\to 0,\Omega=0)$.  For free fermions, it coincides with the density of states.  A product of two Green's functions with close arguments can be represented as a sum of two parts: coherent and incoherent (Ref.~\onlinecite{agd}):
\bwt
\beq
G(\bk+\bq,\omega+\Omega)G(\bk,\omega)=2\pi Z^2\frac{{\bf v}_F\cdot\bq}{{\bf v}^*_F\cdot\bq-i\Omega}\delta(\omega)\delta(\epsilon_k)
+\Phi(\bk,\omega),~\label{g2}
\eeq
\ewt 
where ${\bf v}^*_F=\vf(m/m^*){\hat k}$. A contribution of the coherent part to $\Pi_d^*(\bq\to 0,\Omega=0)$ is equal to the renormalized density of states $\nu Z^2(m^*/m)$. However, it would be incorrect 
 to replace $\Pi_d(\bq\to 0,\Omega=0)$ only by the coherent contribution because there is also a contribution from the incoherent part, $\Phi(\bk,\omega)$, which cannot be evaluated explicitly. A way avoid this complication is
 is to integrate over the momentum first. This integral comes from high energies, of order of the ultraviolet cutoff of the theory, $\Lambda$. 
 At these energies,  one can approximate $G$ by its free-fermion form $G^{-1} (\bk, \omega) = \omega - \epsilon_k$. Integrating over $\epsilon_k$
in between $-\Lambda$ and $\Lambda$ first and then over the frequency, we find that 
$\Pi^*_d(\bq,0)$ is given by the bare density of states:
\bea
\Pi_d^*(q\to 0,\Omega=0)&=&-\nu\int\frac{d\omega}{2\pi}\int^{\Lambda}_{-\Lambda} d\epsilon_k\frac{1}{\left(i\omega-\epsilon_k\right)^2}\notag\\
&&=\nu\int \frac{d\omega}{\pi}\frac{\Lambda}{\omega^2+\Lambda^2}=\nu.
\eea
Both energies ($\omega$ and $\epsilon_k$) in this integral are of order $\Lambda$, which justifies the replacement of the Green's function by its free-fermion form. This result is known in the theory of electron-phonon interaction: 
\cite{phonons} renormalization of the phonon spectrum by particle-hole excitations 
 is the same as if the fermions were free.

Similarly, the $q^2$ term in Eq.~(\ref{pidfree}) is also a high-energy contribution. If the electron spectrum remains quadratic up to $\Lambda$, this term is the same as for free fermions; otherwise, $\kf$ in the prefactor of this term is replaced by some momentum of order of either $\kf$ or of the reciprocal lattice spacing.

 In distinction to the static part of $\Pi_d^*$, the dynamic part  of the bubble--
 the Landau damping term--comes from low-energy fermions by virtue of 
energy conservation. Calculating this part using only the coherent term in Eq.~(\ref{g2}), we see that the Landau damping term is multiplied by an overall factor of $(Z m^*/m)^2$. 
  
The full $\Pi_d^* (\bq,\Omega)$ is then given by
\beq
\Pi_d^* 
(\bq,\Omega) = \nu \left(1 - \frac{q^2}
{b^2} -
 2\cos^2(2\phi_{\bq})\left(\frac{Z m^*}{m}\right)^2 
~\frac{|\Omega|}{\vf q}\right),
\label{x3}
\eeq
 where $b\sim \kf$. 
 
 The $q^2$- and Landau damping
 terms in Eq.~(\ref{x3}) agree with their counterparts in the expression for the polarization bubble in Ref.~\onlinecite{woelfle}. The first, constant term differs, however,  from that in Ref.~\onlinecite{woelfle}, where
 only  the contribution from coherent fermions 
  was included and, as a result,  the first term had an extra $Z$ factor.
 
Substituting Eq.~(\ref{x3}) into Eq. ~(\ref{wed_5}) and 
neglecting again the $q^2/b^2$ term in the bubble compared to the $(aq)^2$ from the interaction, we obtain the dynamic
 effective interaction as
\bwt
\bea
\GRPA(\bk,\omega_k;\bp,\omega_p) &=&\frac{1}{4\nu}
 \frac{d_{\bk}d_{\bp}\ad\bg}{1 + g_{c,2} + |\bk-\bp|^2 a^2
 + 2\cos^2(2\phi_{\bq}) \left(\frac{Z m^*}{m}\right)^2\frac{|\omega_k-\omega_p |}{\vf |\bk-\bp|}}\notag\\
 &&\equiv d_{\bk}d_{\bp}\ad\bg{\bar\Gamma}(\bk-\bp,\omega_k-\omega_p).
\label{t_10a}
\eea 
\ewt
For ${\bk}$ and $\bp$ near the FS, $|\bk -\bp| =2\kf \sin (\theta/2)$
 where $\theta\equiv \angle(\bk,\bp)$. Because $\bk$ and $\bp$ 
 are nearly parallel, 
   $|\bk -\bp| \approx \kf \theta$.
 Notice that by virtue of Eq.~(\ref{su2}), the effective interaction has both charge and spin components even though we neglected the $g_{s,2}$ component of the bare interaction.

A cautionary note: Eq.~(\ref{t_10a}) is obtained within the RPA, 
which does not mix channels with different angular momenta.
 The RPA is valid, strictly speaking, only for a long-range interaction which is not a very realistic assumption.
 For a generic interaction, 
 there is no proof that the effective interaction does not change quantitatively
 beyond the RPA, e.g., it is possible that non-RPA renormalizations destroy a simple pole structure of Eq. (\ref{t_10a}).  In Appendix \ref{app:b},
we discuss the diagrammatic series for $\Gamma^\Omega$ beyond RPA and
 show that the effective interaction does change beyond the RPA.

\subsection{Diagrammatics of the critical FL theory}
\label{sec:3_1}
\subsubsection{Interaction via dynamic collective fluctuations}

\begin{figure}[tbp]
\centering
\includegraphics[width=1.0\columnwidth]{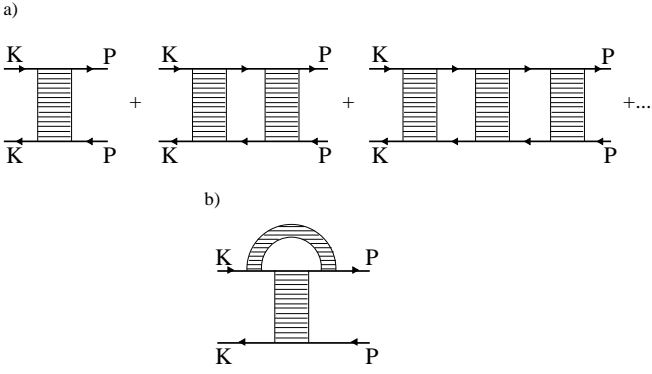}
\caption{ {\emph a}) Ladder series for the full vertex $\Gos$ 
beyond RPA.
 Each hatched block represents $\Gos$ at the RPA level, Eq.~(\ref{t_10a}). 
{\emph b})) One-loop vertex correction to the ladder diagram. 
}  
\label{fig:beyond}
\end{figure}

We now show that the RPA expression for  $\Gos$ [Eq.~(\ref{t_10a})]
is not the full result, even if the control parameter $a\kf$ is large.  \cite{comment_RPA} The reason is that the diagrammatic series for  $\Gos$ in the bare interaction, considered in the previous Section, does not
 include diagrams with particle-hole bubbles at exactly zero momentum and vanishingly small frequency. The argument is that such diagrams vanish due to particle-number conservation.   This is true, however, only if 
 the interaction is static. 
Due to Landau damping, the dressed interaction 
 has a singularity in the complex plane, i,e.,  a branch cut in the Matsubara formalism.
  In this situation,  the frequency 
 integral of the product  $G^2 (q, \Omega)$
 times the dressed interaction has an extra 
  contribution from the
Landau damping singularity.
To account for this effect, we have to re-consider  diagrams with 
 soft particle-hole bubbles and replace each static interaction, $U (\bk-\bp)$, 
by  a dressed one, i.e., by $\Gamma^{\Omega, \mathrm{RPA}}_{\alpha\beta;\gamma\delta}(\bk,\omega_k;\bp,\omega_p)$ given by Eq. (\ref{t_10a}).   Since the 
 RPA
interaction 
 already contains all insertions of particle-hole bubbles,
 the remaining diagrams form the ladder series shown in 
 Fig.~\ref{fig:beyond}. In Secs.~\ref{sec:ladder} and \ref{sec:vertex}  we evaluate 
 this ladder series and 
 show that 
corrections to ladder diagrams are small.

\subsubsection{Renormalization of $\Gos$ beyond RPA}
\label{sec:ladder}

A  building block for ladder diagrams is  the product of 
two Green's functions with the same momentum and frequency and the dynamic interaction (\ref{t_10a}).
We will analyze explicitly the first few terms in the ladder series and then sum up infinite series of diagrams.
Consider first the second-order diagram in  Fig.~\ref{fig:beyond} $a)$. 
 Summing over the internal spin indices and neglecting the momentum transfer $q$ compared to $\kf$ in the $d$-wave formfactors, we obtain
\bwt
\bea
\Gamma^{\Omega,\{2\}}_{\alpha\beta;\gamma\delta}(\bk,\bp)=d_{\bk}d_{\bp}\frac{d_{\bk}^2+d_{\bp}^2}{2}\ad\bg\int \frac{d\Omega}{2\pi}\int \frac{dqq}{2\pi}\int \frac{d\phi}{2\pi}G^2(\bk+\bq,\Omega){\bar\Gamma}(\bq,\Omega){\bar\Gamma}(\bk-\bp+\bq,\Omega),
\label{ladder2}
\eea
\ewt
where $|\bk|=|\bp|=\kf$, $\phi=\angle(\bq,\bk)$, and ${\bar\Gamma}$ is defined in the second line of Eq.~(\ref{t_10a}).  Let us integrate over $\phi$ first. Both the Green's functions and ${\bar \Gamma}$ 
depend on $\phi$. The  $\phi$-dependence of ${\bar \Gamma}$ is in two places: first,
in the anisotropic Landau damping term, 
 which depends on $\phi$ as $\cos^2\left[2(\phi_{\bk}+\phi)\right]$, where $\phi_{\bk}=\angle(\bk,{\hat x})$,
  and, second, in the magnitude of the momentum $|\bk -\bp + \bq|$. We will see, however, that the 
integral over $\phi$ is dominated by narrow regions near  
 $\pm\pi/2$, where $\bq$ is almost perpendicular to $\bk$.
  Therefore,  the prefactor of the Landau damping term can be replaced by
 $\cos^2(2\phi_{\bk})$.
  Expanding next $\phi$ near $\pm \pi/2$ as
$\phi=\pm\pi/2\mp{\bar\phi}$ with $|{\bar\phi}\ll 1$, and using the fact that near a QCP the relevant values of
$|\bk - \bp|\approx \kf\theta$ (determined by the inverse correlation length) are small,  we find that $|\bk-\bp+\bq|^2\approx \left(\kf \theta\pm q\right)^2$ does not depend on ${\bar\phi}$ to leading order. It is thus only $G^2 (\bk + \bq, \Omega)$ that depends on ${\bar\phi}$. The integral over 
 ${\bar \phi}$
  can then be evaluated; one has to be careful though to resolve an ambiguity at $\Omega=0$, which results in a delta-function term: 
 \beq
\int^{\pi}_{-\pi}\frac{d\phi}{2\pi} G^2(\bk+\bq,\Omega)
=Z^2\left(2\frac{\delta(\Omega)}{\vf ^*q}-\frac{|\Omega|}{\left[\Omega^2+(\vf ^*q)^2\right]^{3/2}}\right).
\label{angle}
\eeq
 The delta-function term will give the leading contribution to the vertex. It is
 reproduced if the integral over ${\bar\phi}$ is performed as
 \bea
&&\int^{\pi}_{-\pi}\frac{d\phi}{2\pi} G^2(\bk+\bq,\Omega) \rightarrow  2\int^{\infty}_{-\infty}\frac{d{\bar\phi}}{2\pi} \frac{Z^2}{\left(i\Omega-\vf ^*q{\bar\phi}\right)^2}\nn\\
&=&2i\frac{\partial}{\partial\Omega}\int^{\infty}_{-\infty}\frac{d{\bar\phi}}{2\pi} G(\bk+\bq,\Omega)=2Z^2\frac{\delta(\Omega)}{\vf ^*q}.
\eea
The second term in Eq. (\ref{angle}) is reproduced if one keeps track of the actual limits of the integral over $\phi$. 

Substituting Eq.~(\ref{angle}) into Eq.~(\ref{ladder2}) and integrating over $\Omega$ and $q$, we find, as advertised, that the leading contribution to 
$\Gamma^{\Omega,\{2\}}_{\alpha\beta;\gamma\delta}(\bk,\bp)$ comes from the $\delta-$functional term, while the contribution from the second term in (\ref{angle}) is proportional to $1+ g_{c,2}$, which is small near a QCP. 
The delta-function of $\Omega$ eliminates the frequency dependence of the RPA vertices,
and, integrating over $q$, 
we obtain for the second-order ladder diagram
\bea
\Gamma^{\Omega,\{2\}}_{\alpha\beta;\gamma\delta}(\bk,\bp)&=&
d_{\bk}d_{\bp}\frac{d_{\bk}^2+d_{\bp}^2}{2}\ad\bg
\left(\frac{Z^2m^*\lambda}{2m}\right)\nn\\
&&\times\frac{1}{4\nu} \left[\frac{2}{4(1+\gc)+\left(a\kf \theta\right)^2}\right],
\eea
where 
\beq
\lambda=\frac{1}{2a\kf \sqrt{1+\gc}}.
\label{lambda}
\eeq
We will see below that  $Z^2 m^* \lambda/m \approx 1$
 in the critical FL regime. This implies $\Gamma^{\Omega,\{2\}}$ is of the same order as $\Gamma^{\Omega, \mathrm{RPA}}$ and, thus, ladder renormalizations cannot be neglected. We  also note that the parameter $\lambda$ will play the key role in our further analysis. In particular, it will be shown
 that   the critical FL regime is defined by a condition $\lambda \ll 1$ rather than by $1 + g_{c,2} \ll 1$. Because $\lambda$ contains a small factor of $1/a\kf\ll 1$, the former is stronger than the latter.  
 
We now return to the ladder series composed from the $G^2 (q, \Omega)$ blocks. 
Repeating the same procedure for the $n$th term of the series, we obtain
\bea
\Gamma^{\Omega,\{n\}}_{\alpha\beta;\gamma\delta}(\bk,\bp)&=&
d_{\bk}d_{\bp}\frac{d_{\bk}^{2(n-1)}+d_{\bp}^{2(n-1)}}{2}\ad\bg\nn\\
&&\times\left(\frac{Z^2m^*\lambda}{2m}\right)^{n-1}
{\bar\Gamma^{\{n\}}}(\theta),
\eea
where
\beq
{\bar\Gamma^{\{n\}}}(\theta)=\frac{1}{4\nu} \left[\frac{n}{n^2(1+\gc)+\left(a\kf \theta\right)^2}\right].
\label{gn}
\eeq
Comparing Eq.~(\ref{gn}) to Eq.~(\ref{t_10a}), we see that the structure of the RPA interaction (the $n=1$)
 term is not reproduced at higher orders: first, the dependence on the angle $\theta$ between $\bk$ and $\bp$ is different and, second, there are additional $d$ factors arising at each order of the perturbation theory. However, the angular dependence can be scaled back to that of the $n=1$ term by replacing $\theta\to n\theta$. This means that  the angular harmonics 
$\int d\theta  {\bar\Gamma^{\{n\}}}(\theta) \cos(\ell\theta)/2\pi$ of ${\bar\Gamma^{\{n\}}}$ are the same as those of ${\bar\Gamma}$, as long as  $\ell$ and $n$ are not too large: $\ell n\ll \lambda$. 
Since $\lambda \ll 1$  in the critical FL regime, this condition is always satisfied for 
not too large $l$ and $n$. In other words, ${\bar\Gamma^{\{n\}}}$ can be re-written as the sum of two terms: the first one is the same as for $n=1$ and the second one vanishes on angular integration
\beq
{\bar\Gamma^{\{n\}}}(\theta)={\bar\Gamma}(\theta)+R^{\{n\}}(\theta),
\eeq
where
\beq
R^{\{n\}}(\theta)=(n-1)\frac{(a\kf\theta)^2-n(1+\gc)}{\left[n^2(1+\gc)+(a\kf\theta)^2\right]\left[1+\gc+(a\kf\theta)^2\right]}
\eeq
and $\int d\theta R^{\{n\}}=0$.
Since the observables are determined by harmonics of $\Gamma^\Omega$, we will neglect the remainder term $R^{\{n\}}$ below, i.e, replace
${\bar\Gamma^{\{n\}}}(\theta)$ by ${\bar\Gamma}(\theta)$.

Still, the subsequent terms in the series contain additional factors
$d^{2(n-1)}_\bk + d^{2(n-1)}_\bp$. These terms account for renormalization of a 
smooth, regular variation of $\Gamma^{\Omega} $ along the FS.  Such a variation,  inherent to a model with
anisotropic, $d$-wave, $\cos(2\phi)$ interaction,  implies that both the
quasiparticle residue and effective mass are not uniform along the FS. The $d$-channel vertex can be obtained 
by extracting contributions proportional to $d_{\bk} d_{\bp} $  at any order, neglecting all other terms,
 and summing up the series.   
On the other hand, one can average the  vertex over the FS, i.e.,  restore Galilean invariance broken by neglecting the $\sin(2\phi)$ interaction.
The two approaches lead to physically equivalent results which differ only by  numerical prefactors.
 For brevity, we only discuss here 
 the approach based on averaging, which is 
 particularly appealing because it allows one to use the  technique developed for isotropic FL systems to obtain $m^{*}/m$ and $Z$. 
 Results of 
 alternative 
  approaches are presented in Appendix~\ref{app:a}. 
We emphasize that, although no important physics is lost in either of 
the approaches, all of them are still approximate.  
We will discuss one subtle issue associated with this approximation in
Appendix.~\ref{sec:subtle}. 

We now apply the averaging procedure.  Since relevant $\bk$ and $\bp$ are almost parallel to each other, we will set $\bk=\bp$ in the $d$ factors and use an identity
\beq
\langle d^{2n}_{\bk}\rangle=\frac{(2n-1)!!}{n!}.
\eeq
Substituting this expression into  $\Gamma^{\Omega, \{n\}}$ and summing up the series, we find
 that the ladder series for $\Gos$ result
 in overall renormalization of $\GRPA$
\beq
\Gos= \frac{1}{Z_{\Gamma}}\GRPA = 
\frac{1}{Z_{\Gamma}}\frac{1}{4\nu}\frac{\delta_{\alpha \delta} \delta_{\beta \gamma}}{1+\gc+(a\kf \theta)^2},
\label{gfull}
\eeq
where
\bea
\frac{1}{Z_{\Gamma}}&=&\sum_{n=1}^{\infty}\left(\frac{Z^2m^*}{2m}\lambda\right)^{n-1}\frac{(2n-1)!!}{n!}\nn\\
&&=\frac{2}{\sqrt{1-\frac{Z^2m*}{m}\lambda}\left(1+\sqrt{1-\frac{Z^2m*}{m}\lambda}\right)}.
\label{zgamma}
\eea
Since $Z_\Gamma\leq 1$, ladder renormalization enhances the interaction compared to the RPA result.
\subsubsection{Vertex corrections}
\label{sec:vertex}

We now show that vertex corrections to ladder series are small and thus Eq.~(\ref{gfull}) is a complete result for our model. The lowest order vertex correction to $\Gamma^{\Omega}$ is shown in Fig.~\ref{fig:beyond}$b$.
 When calculating the ladder diagram in the previous Section, we saw that the interaction vertices
  are effectively static ($\Omega=0$), while typical bosonic momenta $q\sim \sqrt{1+\gc}/a$ are small near a QCP. 
 To estimate the vertex correction, we can then 
 simply calculate a three-leg vertex, shown in Fig.~\ref{fig:beyond}$c$, 
 at zero external frequency and finite but small external momentum $q$.
 With these simplifications,  the three-leg vertex reduces to
\bea
\Gamma_{\triangle}(\bq\to 0,\Omega=0)&=&\int \frac{d\Omega'}{2\pi}\frac{d^2q'}{(2\pi)^2}G(\bk+\bq',\Omega')\nn\\
&&\times G(\bk+\bq'+\bq, \Omega'){\bar\Gamma}(\bq',\Omega').\nn\\
\label{triangle}
\eea
We 
 neglected
  the  spin and $d$-wave factors as well as the anisotropy of the Landau damping term, all of which give only overall numerical prefactors. 
 Since ${\bar\Gamma}(q',\Omega')$ is 
 now
 isotropic, the angular integration in Eq.~(\ref{triangle}) can be performed first.  Because the momenta in two Green's functions differ by small yet finite $q$, the angular integral of $G(\bk+\bq',\Omega')G(\bk+\bq'+\bq,\Omega')$
contains only a regular but no $\delta(\Omega')$ term, i.e., the result is given by the second term in Eq.~(\ref{angle}). Hence
\bea
\Gamma_{\triangle}(\bq\to 0,\Omega=0)&=&-Z^2\int \frac{d\Omega'}{2\pi}\frac{dq'q'}{2\pi}\frac{|\Omega'|}{\left[(\Omega')^2+(\vf^*q')^2\right]^{3/2}}\nn\\
&&\times{\bar\Gamma}(\bq',\Omega').
\label{triangle_1}
\eea
We notice immediately that the remaining double integral is finite even right at a QCP, where $1+\gc=0$. This is to be contrasted with ladder diagrams of the previous section 
 which 
 diverge near a QCP.
 An explicit calculation can be carried out by rescaling the variables as $q'=x q_0 $ and $\Omega'= \Omega_0 y$, where
 \beq
q_0=Z\sqrt{m^*/m}/a,\;\Omega_0=q_0\vf^*,
\label{q0omega0}
\eeq
 and introducing polar coordinates $x=r\cos\psi$, $y=r\sin\psi$.
We then obtain 
\bea
&&\Gamma_{\triangle}(\bq\to 0,\Omega=0) = - Z \left(\frac{m^*}{m}\right)^{1/2}\frac{1}{a\kf} \times \nonumber \\
&& \frac{1}{4\pi}\int_0^\infty dr \int_0^{\pi/2} d \psi \frac{\sin{\psi} \cos{\psi}}{r^2 \cos^2{\psi} + \alpha + \tan{\psi}},
\eea
where
\beq
\alpha=(1+\gc) \frac{m}{m^*Z^2}.
\label{alpha}
\eeq
 Integrating over $r$, we obtain
\beq
\Gamma_{\triangle}(\bq\to 0,\Omega=0)=-\frac{Z}{8}\left(\frac{m^*}{m}\right)^{1/2}\frac{1}{a\kf}{\cal F}(\alpha),
\label{ch_3}
\eeq
where
 \beq
{\cal F}(x)=\int^{\pi/2}_0 d\psi \frac{\sin{\psi}}{\sqrt{x+\tan\psi}}.
\label{alpha1}
\eeq
For $x\ll 1$,
\bea
{\cal F}(x)&=&\frac{\Gamma^2(3/4)}{\sqrt{\pi}}-\frac{\Gamma^2(1/4)}{8\sqrt{\pi}}x+\dots\nn\\
&&\approx 0.847-0.927 x +\dots
\eea
In the next Section, we show that $\alpha \sim 1$ away from a QCP
 and $\alpha\ll 1$ near a QCP. In both cases, ${\cal F} (x)
  \sim 1$.
Therefore, the vertex correction can be 
estimated as
 \beq
 \Gamma_{\triangle}(\bq\to 0,\Omega=0)\sim Z\left(\frac{m^*}{m}\right)^{1/2}\frac{1}{a\kf},
 \label{gamma_three_leg}
 \eeq
 which is at most of order $1/a \kf \ll 1$.

 Equation (\ref{gamma_three_leg}) can be re-written  
via the characteristic momentum $q_0$ or characteristic frequency $\Omega_0$ from Eq.~(\ref{q0omega0}) as 
\beq
\Gamma_{\triangle}(\bq\to 0,\Omega=0)\sim q_0/\kf \sim \Omega_0/\ef^*,
\label{est}
\eeq
where $E_{\mathrm{F}}^*=\kf\vf^*$.
When written in this form, 
it is clear that 
$\Gamma_{\triangle}(\bq\to 0,\Omega=0)$ is 
an effective Migdal (adiabatic) parameter, i.e., the ratio of characteristic energies of bosons $(\Omega_0$) and fermions ($\ef^*$), of our theory.\cite{rech}
For $a\kf \gg 1$, this ratio is small both away and near a QCP.
 The Migdal parameter occurs here in same way as for the case of electrons interacting with  {\it optical} phonons. Namely, if the interaction in
Eq.~(\ref{triangle}) is replaced by a propagator of optical phonons $\Omega_0^2/(\Omega^2+\Omega_0^2)$ (where now $\Omega_0\ll \ef$ is the optical phonon frequency), the result for the vertex is the same (up to a number). The reason for the smallness of the vertex in both cases is a small phase space defined by $q_0$ and $\Omega_0$. Although the vertex correction is also small for {\it acoustic} phonons, the reason  is different in this case: a typical momentum of acoustic phonon is of the order of the inverse lattice spacing ($\sim \kf$), while the electron energy at the same momentum is of the order of $\ef$, i.e., much larger than the phonon energy $\Omega_0$.   

Two cautionary notes, both related to the fact that 
 the integral in Eq.~(\ref{ch_3}) 
is determined by 
$\Omega' \sim \vf^* q' \sim (\ef/(a\kf) Z\sqrt{m/m^*}$.
First, we used a quasistatic, $\Omega'/\vf^* q'$ form of 
the Landau damping term, which is not, strictly speaking, justified in this regime. However, because $q'$ is not much smaller than $\Omega'/\vf^*$ either, the result for $\Gamma_{\triangle}$ is still correct up to a numerical prefactor. Second, and more important, 
 we  assumed a 
  renormalized 
 FL  form of the Green's function
  $G(\bk, \omega) = Z/\left[\omega - \vf^* (k-\kf)\right]$.  Meanwhile,
 we will see later
 that (i) a FL behavior  extends only 
up to frequencies of order $\omega_{FL} \sim  (\vf/a) (1+ g_{c,2})^{3/2}$, and (ii) 
 $\Omega_0\sim Z\sqrt{m/m^*}(\vf^*/a) \sim (\vf/a) (a\kf)^{3/2} (1 + g_{c,2})^{3/4}$. 
 Thus, typical $\Omega'\sim\Omega_0\ll\omega_{\mathrm{FL}}$ are outside the FL regime  for small enough $1 + g_{c,2}$,. 
 However, the non-FL effects are not important here either
 because the three-leg vertex outside the FL regime 
is determined by very high energies ($\sim v_F/a$), where fermions behave as almost free quasiparticles with $Z,m^*/m\approx 1$ (Ref.~\onlinecite{rech}).
 The result is that Eq.~(\ref{est}) still holds, but the
relevant fermionic energy is $\ef$ while the relevant bosonic energy is $\Omega_0 \sim \ef/a\kf$, and the vertex correction is simply given by 
\beq
\Gamma_{\triangle}(\bq\to 0,\Omega=0) \sim 1/a\kf.
\eeq  Certainly, the same result for the vertex is obtained if one substitutes the free Green's function and bare vertex into  Eq.~(\ref{gamma_three_leg}) from the very beginning.  We also emphasize that the vertex correction was found to be small in the limit $\Omega/q \to 0$. In the opposite limit, $q/\Omega \to 0$, vertex corrections are not small and  must be included into the theory to ensure that the Ward identities are satisfied. 

\subsubsection{Complete form of the FL vertex}

To summarize this Section, we conclude that, in a critical FL, the full vertex 
 $\Gos$ 
differs from the RPA result by a constant renormalization factor $1/Z_{\Gamma}$ given by Eq.~(\ref{zgamma}).
Hence a complete result for the static 
 $d-$wave vertex, averaged over the FS, and for $\bk\approx\bp$ reads 
\beq
\Gos (\bk,\bp) 
= \frac{1}{4\nu Z_\Gamma} \frac{\delta_{\alpha \delta} \delta_{\beta \gamma}}{1 + g_{c,2} + (a\kf\theta)^2}.
\label{wed_66}
\eeq
 To obtain an explicit form of 
$\Gos$, we need to know $Z$ and $m^*/m$.
This is what we discuss in the next Section.
We show there that, near the QCP, the leading terms in $Z$ and $m^*/m$ are
determined  
by the static vertex,
i.e., there is no need to invoke  
dynamics of bosonic modes.
The subleading terms, however, 
require the knowledge of the dynamic vertex. 
Following the same steps that
led us to 
Eq.~(\ref{wed_66}),
one can show that, for $a\kf\theta\gg\sqrt{1+g_{c,2}}$, the dynamic vertex
is given by
\bwt
\beq
\Gos (\bk,\omega_k;\bp,\omega_p) 
= \frac{1}{4\nu Z_\Gamma} \frac{\delta_{\alpha \delta} \delta_{\beta \gamma}}{1 + g_{c,2} + (a\kf\theta)^2
+{\cal C}\left(\frac{Z m^*}{m}\right)^2~\frac{|\omega_k-\omega_p|}{\vf\kf \theta}},
\label{wed_666}
\eeq
\ewt  
where ${\cal C}\sim 1$.  
In a general case,  the dynamic vertex is a scaling function of the following form
\bwt
\beq
\Gos (\bk,\omega_k;\bp,\omega_p) 
= \frac{1}{4\nu Z_\Gamma} \frac{\delta_{\alpha \delta} \delta_{\beta \gamma}}{1+g_{c,2}}
{\cal G}\left(\frac{a\kf\theta}{\sqrt{1+g_{c,2}}},\left(\frac{Zm^*}{m}\right)^2\frac{a|\omega_k-\omega_p|}{v_F(1+g_{c,2})^{3/2}}\right),
\eeq
\ewt
where ${\cal G}(x,0)=1/(1+x^2)$. As an explicit form of ${\cal G}$ will not be required for the estimates of the subleading terms in $Z$ and $m^*/m$,
we will use
below the approximate Eq.~(\ref{wed_666})  
with ${\cal C}=1$.

\section{Pitaevski-Landau relations  and self-consistent equations for $Z$ and $m^*/m$}
\label{sec:4}
\subsection{Pitaevski-Landau identities for the derivatives of the self-energy}
The renormalized mass can be found in two ways. If the Landau function is known, $m^*/m$ can be determined from its $g_{c,1}$ component.  Alternatively, one can compute the self-energy in a given microscopic model
and extract 
 mass renormalization from the renormalized Green's function.   In general, the quasiparticle residue $Z$ can be found only from a microscopic calculation because renormalization of $Z$ is determined by states away from the FS,  which are not described by the FL theory. Renormalizations of $Z$ and $m^*/m$ near QCPs have
 been considered in \cite{Lee,ccm,monien,Khvesh,aim,Oganesyan,Kee,MRA,cmgg,chub_cross,rech,cgy,CK06,Lawler,senthil,cm_chi,tigran,khodel}
 via loop-wise expansions in the effective interaction. Near a QCP, however,
 typical  energies involved in renormalizations 
 vanishe in the inverse proportion to the divergent correlation length.
    This allows one to determine both $m^*/m$ and $Z$ within the FL framework. To do so, 
 we will use the Pitaevski-Landau identities which relate the derivatives of the self-energy \cite{agd,lp} to $\Gamma^\Omega$ from (\ref{wed_66}).
 
 We will need three of the Pitaevski-Landau identities
\bea
&&  \frac{\partial \Sigma}{\partial \omega} = - i \sum_\beta \int \Gamma^\Omega_{\alpha \beta, \alpha \beta} (K_F, P) \left[G^2 (P)\right]_\Omega \frac{ d^3 P}{(2\pi)^3} \nonumber \\
&& \frac{\partial \Sigma}{\partial \bk} = 
 i \sum_\beta \int  \frac{\bp}{m} \Gamma^q_{\alpha \beta, \alpha \beta} (K_F, P)
\left[G^2 (P)\right]_q \frac{ d^3 P}{(2\pi)^3} \nonumber \\
&&  \bk \frac{\partial \Sigma}{\partial \omega} =  - i \sum_\beta \int
\bp  \Gamma^\Omega_{\alpha \beta, \alpha \beta} (K_F, P) 
\left[G^2 (P)\right]_\Omega \frac{ d^3 P}{(2\pi)^3}. \nonumber \\
&& 
\label{tu_1}
\eea
Here, as before, a shorthand $K_F$ denotes the \lq\lq four-momentum\rq\rq~ on the FS, i.e. $K_F\equiv\{\kf{\hat k},\omega_k=0\}$, while the objects $\left[G^2(P)\right]_\Omega$ and $\left[G^2(P)\right]_q$  represent 
the product $G(P) G(P+Q)$ taken at vanishing $Q$ 
in the limits of $q/\Omega\to 0$  and $\Omega/q \to 0$,
 respectively.  Notice that the internal momentum $P$ is generally not at the FS. Also, we switched here from the  Matsubara formalism, employed in the previous sections, to the causal one. To simplify notations, we use the same symbols for Matsubara and real frequencies in all cases when it does not lead to a confusion.

The first two relations are essentially the Ward identities following from particle-number conservation (gauge invariance), while  the third one is special for Galilean-invariant system. We remind that Galilean invariance was restored in our problem
by averaging over the FS.  The vertices  $\Gamma^\Omega$ and $\Gamma^{q}$ are related by Eq. (\ref{qomega}). 

Equations (\ref{tu_1}) and (\ref{qomega}) determine a Taylor expansion of the self-energy $\Sigma (k, \omega)$ to first order in $\omega$ and $\epsilon_k^* = \vf^*  (k-\kf )$:
\begin{widetext}
\beq
\Sigma (K) = i\sum_\beta \left[-(\omega - \epsilon_k)\int \Gamma^\Omega_{\alpha \beta, \alpha \beta} (K_F, P) \left[G^2(P)\right]_\Omega \frac{d^3 P}{(2\pi)^3}
 + 
 \frac{\epsilon^*_k}{Z}   \int \Gamma^\Omega_{\alpha \beta, \alpha \beta} (K_F, P) \left[\left[G^2(P)\right]_q - \left[G^2(P)\right]_\Omega\right] \frac{\bp\cdot\bk_\mathrm{F}}{\kf^2} \frac{d^3 P}{(2\pi)^3}\right],
\label{wed_1}
\eeq
\end{widetext}
 where the FS term $\Sigma(K_F)$ is absorbed into a shift of the chemical potential.
 Using an identity
\beq
\left[G^2(K)\right]_q - \left[G^2(K)\right]_\Omega = - \frac{2 \pi i Z^2 m^*}{\kf } \delta (\omega) \delta(k - \kf ),
\label{wed_2}
\eeq
valid for any order of integration over the fermionic momentum and frequency, we can rewrite  Eq.~(\ref{wed_1}) as
\begin{widetext}
\bea
\Sigma_(K) = \left(\omega - \epsilon_k\right) 
 \left[-i \sum_\beta \int \Gamma^\Omega_{\alpha \beta, \alpha \beta} (K_F, P) \left[G^2 (P)\right]_\Omega \frac{ d^3 P}{(2\pi)^3}\right] +  \epsilon_k \frac{Z m}{4 \pi^2}
 \sum_\beta \int \Gamma^\Omega_{\alpha \beta, \alpha \beta}
 (\theta) \cos \theta d \theta.
\label{wed_2a}
\eea
\end{widetext}
Equivalently,
\beq
\Sigma (K) = 
(\omega - \epsilon_k) \left[\frac{1}{Z} -1\right] + \frac{\epsilon_k}{Z} \left[1 - \frac{m}{m^*}\right],
\label{wed_3}
\eeq
where
\beq
\frac{1}{Z} = 1 -i \sum_\beta \int \Gamma^\Omega_{\alpha \beta, \alpha \beta} (K_F, P) \left[G^2(P)\right]_\Omega \frac{d^3 P}{(2\pi)^3} 
\label{wed_4a}
\eeq
\beq
\frac{1}{m^*} = \frac{1}{m} - \frac{Z^2}{4 \pi^2}
 \sum_\beta \int \Gamma^\Omega_{\alpha \beta, \alpha \beta}
 (\theta) {\cos \theta} d \theta,
\label{wed_4b}
\eeq
and $\theta$ is the angle between $\bk$ and $\bp$ when both momenta are on the FS. 

The integral in Eq.~(\ref{wed_4b})  involves
 a full {\it static} vertex of the interaction between the particles on the FS. In our case, this vertex is given by 
Eq.~(\ref{gfull}) with $\omega_k=\omega_p$. Using this equation, we find
\beq
 \frac{Z^2}{4 \pi^2}
 \sum_\beta \int \Gamma^\Omega_{\alpha \beta, \alpha \beta}
 (\theta) {\cos \theta} d \theta = \frac{Z^2}{4Z_\Gamma} \lambda
\label{wed_4c}
\eeq
and
\beq
\frac{m^*}{m} = 1 +
\frac{\lambda}{4}\frac{Z^2}{Z_{\Gamma}}\frac{m^*}{m}.
\label{x1}
\eeq
This is an implicit equation for 
$m^*/m$. 

Next, we consider the $Z$ factor. 
In contrast to the effective mass, the $Z$ factor is, in general, determined by the dynamic vertex. However, we will see shortly
that the leading term in $Z$ still comes from the static vertex. 
 To see this, we substitute the dynamic interaction from Eq.~(\ref{wed_666}) into Eq.~(\ref{wed_4a}) and  
re-label 
 the variables as $\bq=\bp-\bk_{\mathrm{F}}$ and $\Omega
=\omega_p$, upon which Eq.~(\ref{wed_4a}) reduces to
 \begin{widetext}
\beq
\frac{1}{Z} = 1 - \frac{i Z^2}{4 Z_\Gamma \nu} \int \frac{d^2 q d \Omega}{(2\pi)^3} 
\frac{1}{(\Omega+{\tilde\Omega}  - \epsilon^*_{\bk_{F} + \bq}+i \delta_\epsilon)^2}\Big\vert_{{\tilde\Omega}\to 0}
~\frac{1}{1 + g_{c,2} + (a q)^2 - \left(\frac{Z m^*}{m}\right)^2~ 
i \frac{|\Omega|}{\vf q}}.
\label{wed_7}
\eeq
\end{widetext}
This integral is similar to that in Eq.~(\ref{ladder2}) for the second term in the ladder series for $\Gos$,
 except for now we have one rather than two interaction vertices. The result of the angular integration is the same as in Eq.~(\ref{angle}).
 The leading contribution to $1/Z$ comes from the $\delta$-function term in Eq.~(\ref{angle}), which renders the vertex static. 
  The contribution from the second, 
  dynamic
   term Eq.~(\ref{angle}) is proportional to the three-leg vertex, estimated in Sec. ~\ref{sec:vertex}, except for an extra factor of $1/Z_{\Gamma}$ because the $Z$ factor contains a full rather than RPA vertex.  
   Since the three-leg vertex is small, so is the dynamic correction to $1/Z$. 
   Collecting all terms, we find
\beq
\frac{1}{Z} = 1 + \frac{\lambda}{4} \frac{Z^2}{Z_{\Gamma}}\frac{m^*}{m} 
- \frac{m^*}{m} \Gamma_\Delta,
\label{wed_8a}
\eeq
where $\Gamma_\Delta \sim \Gamma_\triangle \sim 1/a\kf\ll 1$
is the dynamic correction due to a three-leg vertex.

\subsection{Solutions for $m^*/m$, $Z$, and  $Z_\Gamma$ near a QCP}

Equations (\ref{x1}), (\ref{wed_8a}), and (\ref{zgamma}) form a closed set from which one obtains $m^*/m$, $Z$ and $Z_\Gamma$ as functions of the coupling constant $\lambda$, defined by Eq.~(\ref{lambda}).  
Neglecting $\Delta$ in  Eq.~(\ref{wed_8a}) and comparing it to Eq.~(\ref{x1}) and, we see that 
\beq
Z \frac{m^*}{m}=1.
\label{y_8}
\eeq
 Using this relation, we can re-write  Eqs.~(\ref{wed_8a}) and 
(\ref{zgamma}) as
 \bea
&& \frac{1}{Z_\Gamma} = \frac{4}{Z} \left(\frac{1-Z}{\lambda Z}\right) \label{x11aa} \\
&&\frac{1}{Z}=1+\frac{1}{2}\frac{\lambda Z}{\sqrt{1-\lambda Z}\left(1+\sqrt{1-\lambda Z}\right)}.
\label{x11a}
\eea
Introducing a new variable $x=\sqrt{1-\lambda Z}$, we re-write 
 Eq. (\ref{x11a})  as a cubic equation for $x$:
 \beq
x^3+x^2+(2\lambda-1)x=1.
\label{y_9}
\eeq
 The only real solution of this equation is 
\beq
x=\frac{1}{3}L-\frac{2(\lambda-2/3)}{L}-1/3,
\eeq
where $L=\left(9\lambda +8+3\sqrt{3\lambda}\sqrt{8\lambda^2-13\lambda+16}\right)^{1/3}$. Once $x$ is known,
the three parameters of the FL theory $m^*/m$, $Z$, and $Z_{\Gamma}$ are also known.
For $\lambda\ll 1$, $x=1-\lambda/2+{\cal O}(\lambda^3)$; for $\lambda\gg 1$, $x=1/2\lambda+1/4\lambda^2+{\cal O}(1/\lambda^4)$. Using these asymptotics, we immediately find that at weak coupling
\beq
\frac{m^*}{m}\big\vert_{\lambda\ll 1}= 1+\lambda/4;~ Z\big\vert_{\lambda\ll 1}= 1-\lambda/4;~Z_{\Gamma}\big\vert_{\lambda\ll 1} =1-3\lambda/4,
\label{mz_1}
\eeq
while at strong coupling
\bea
&&\frac{m^*}{m}\big\vert_{\lambda\gg 1}=\lambda+1/4\lambda;~;Z\big\vert_{\lambda\gg 1}=1/\lambda-1/4\lambda^3;~ \nonumber \\
&& Z_{\Gamma}\big\vert_{\lambda\gg 1}=1/4\lambda+1/4\lambda^2.
\label{mz}
\eea
Notice that $Z_{\Gamma} \approx Z/4$ in the strong coupling limit. 

We also emphasize that our calculations show that the efective mass $m^*$ diverges right at the critical point but not earlier. In this respect, our results agree with Ref.~\onlinecite{cgy} but 
 disagree with Ref.~\onlinecite{khodel}, where it was conjectured that the effective mass
  may
  diverge
   at a 
  topological transition which pre-empts a QCP.
 Such a behavior would follow from our  formalism only
 if, for some reason, renormalization of $Z$ and $Z_\Gamma$ occurrs in such a way that $C_Z=Z^2/Z_{\Gamma}$ were an invariant. According to Eq.~(\ref{x1}), the effective mass $m^*/m=(1-\lambda C_Z/4)$ would then diverge at finite $\lambda=4/C_Z$.
However,  
 the solution of the full set of equations 
shows that the invariant is $Z/Z_{\Gamma}$, in which case the divergence of $m^*$ is only possible at $\lambda=\infty$.

\subsection{Frequency and momentum dependences of the self-energy}
\label{sec:fullsigma}

We now return to the self-energy given by Eq.~(\ref{wed_3}).
In the previous section, we found that the relation $Zm^*/m=1$ holds as long as the vertex correction $\Delta$
 is small.
Substituting this relation 
into Eq.~(\ref{wed_3}),  we find that  $\Sigma$ is \lq\lq local\rq\rq\--
 it  depends  on $\omega$ but not on $\epsilon_k$:
\beq
\Sigma (K) \approx (m^*/m-1)\omega\approx \lambda\omega.
\label{wed_17a}
\eeq
Introducing the  correlation length $\xi=a/\sqrt{1+g_{c,2}}$, we can re-write the mass renormalization coefficient as $\lambda=\xi/\kf a^2$. Notice that $\xi$ is the bare correlation length which enters the RPA formula for the vertex.  Rosch and W{\"o}lfle \cite{woelfle} obtained the same expression for  $m^*/m$ in terms of  $\xi$  but, for reasons displayed in Sec.~\ref{sec:away}, their $\xi$ contains an additional factor of $1/\sqrt{Z}$ compared to ours.

The $\epsilon_k$-dependence of $\Sigma$ is determined by the vertex correction $\Delta$.
 Substituting a full form of of $1/Z$ from Eq.~(\ref{wed_8a})
 into Eq.~(\ref{wed_3}),
  we find
\beq
\Sigma (0,\epsilon_k,) = \epsilon_k\left( 1 - \frac{m}{m^*Z} \right) = \Gamma_{\triangle} \epsilon_k \sim \frac{\epsilon_k}{a\kf}. 
\label{wed_17}
\eeq
We see that  the k-dependent part of the self-energy remains small near a QCP.

\section{Critical FL theory: Landau parameters}
\label{sec:3}

Explicit solutions for $m^*$, $Z$, and $Z_{\Gamma}$,
 obtained in the previous Section, complete our task of obtaining the FL vertex: one just needs to substitute these solutions into Eq.~(\ref{wed_66}). Having found the full vertex, we can now construct the FL theory of the critical region, i.e.,  relate harmonics of $ \Gamma^\Omega_{\alpha\beta;\gamma\delta}$ to observables.
 We emphasize again that $\Gos$ plays the role of an effective interaction between quasiparticles of an ordinary FL 
with \lq\lq bare\rq\rq\ nematic susceptibility  $\chi_{c,2} \propto 1/(1 + g_{c,2})$ already enhanced by the interactions. We label \lq\lq bare\rq\rq\ susceptibilities as $\chi_{a,n}$ and 
 the ones renormalized by $\Gos$ as ${\bar \chi}_{a,n}$.
Applying the standard FL phenomenology, we
 obtain 
\bea
&&\frac{m^*}{m} = 1 + {\bar g}_{c1} \nonumber \\
&&
{\bar  \chi}_{c,n} = \chi_{c,n} \frac{1+ {\bar g}_{c1}}{1 + {\bar g}_{c,n}},~~~
 {\bar \chi}_{s,n} = \chi_{s,n} \frac{1+ {\bar g}_{c1}}{1 + {\bar g}_{s,n}},  
\label{t_11}
\eea
where ${\bar g}_{c,n}$ and ${\bar g}_{s,n}$ are charge and spin components  of the Landau function for the critical FL 
\begin{widetext}
\beq
{\bar g}_{\alpha\beta;\gamma\delta} (\theta) = 2 \nu Z^2 \frac{m^*}{m} 
\Gamma^\Omega_{\alpha\beta;\gamma\delta} (\theta)  =  \frac{Z^2m^*}{4 Z_{\Gamma}m} \frac{1}{1 + g_{c,2} + (a\kf\theta)^2}
\left(\ag\bd+{\vec\sigma}_{\alpha\gamma}\cdot{\vec\sigma}_{\beta\delta}\right) \approx
 \frac{1}{1 + g_{c,2} + (a\kf\theta)^2}
\left(\ag\bd+{\vec\sigma}_{\alpha\gamma}\cdot{\vec\sigma}_{\beta\delta}\right).
\eeq
\end{widetext}

In 2D,
the harmonics of
${\bar g} (\theta)$ are given by
${\bar g}_n = \oint {\bar g} (\theta) \cos ({n\theta}) d\theta/(2\pi)$.
Integrating over $\theta$,
 we find that the first $n\lesssim\lambda$ harmonics of ${\bar g}$ both in the charge and spin channels diverge
concurrently with the effective mass upon approaching QCP as
\beq
{\bar g}_{a,n} = \lambda,
\label{y_10}
\eeq
 where $\lambda$ is given by (\ref{lambda}). At a QCP, where $\lambda\to \infty$, {\it all} harmonics of ${\bar g}$
 diverge.
 
We also see, however,
 the susceptibilities retain their bare values
  despite the divergence of the Landau components.  Indeed, because all Landau components diverge in the same way, renormalization of the effective mass in the numerator cancels with that of the effective \lq\lq$g$-factors\rq\rq\ :
 \beq
{\bar \chi}_{a,n} = \chi_{a,n} \frac{1 + {\bar g}_{c,1}}{1 + {\bar g}_{a,n}} 
 \approx \chi_{a,n} \frac{{\bar g}_{c,1}}{{\bar g}_{a,n}} \approx  \chi_{a,n}
\label{dm_2}
\eeq
In particular, the nematic susceptibility remains
equal to $\chi_{c,2} \propto 1/(1 + g_{c,2})$, 
i.e., {\it it is not affected by mass renormalization}. 

\begin{figure}[tbp]
\centering
\includegraphics[width=0.9\columnwidth]{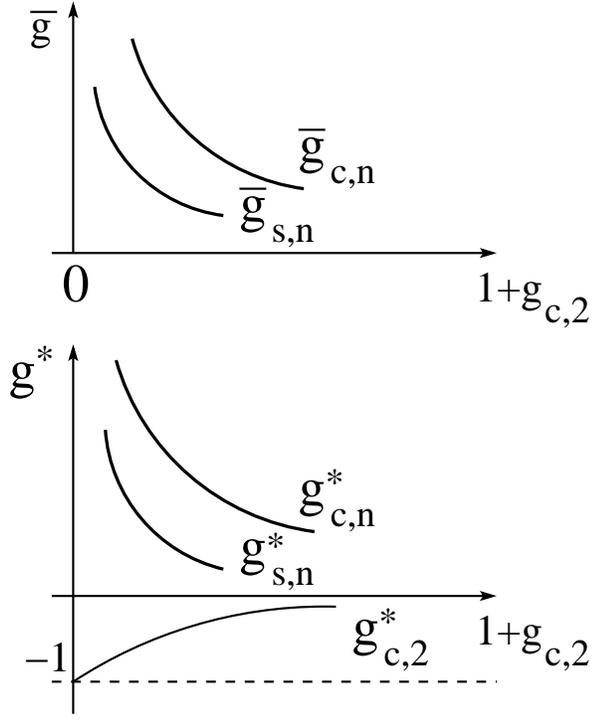}
\caption{Top: schematic behavior of the charge ($c$) and spin ($s$) components of the critical Landau function ${\bar g}$ as a function of $g_{c,2}$. Bottom: same for the actual Landau function $g^*$. }
\label{fig:sigma_1}
\end{figure}

One can also introduce 
 an
\lq\lq actual\rq\rq\/ Landau function, $g^*$, which describes a combined effect of renormalizations in the ordinary and critical FLs and includes contributions from both the regular and  collective mode parts of $\Gos$. 
The nematic susceptibility can be written equivalently either 
  in terms of either ${\bar g}$ or  $g^*$:
\beq
 {\bar \chi}_{c,2} = \chi_{c,2}  \frac{1 + {\bar g}_{c,1}}{1 + {\bar g}_{c,2}} =  \frac{\chi^{\{0\}}_{c,2}}{1 + g_{c,2}} \frac{1 + {\bar g}_{c,1}}{1 + {\bar g}_{c,2}}
 =\chi^{\{0\}}_{c,2} \frac{1 + g^*_{c,1}}{1 + g^*_{c,2}}.
\label{x2}
\eeq
All other susceptibilities $ {\bar \chi}_{a,n}$ with $\{a,n\}$ different from $\{c,2\}$ can be written as 
\beq
{\bar \chi}_{a,n} = \chi^{\{0\}}_{a,n} \frac{1 + {\bar g}_{c,1}}{1 + {\bar g}_{a,n}}
 =\chi^{\{0\}}_{a,n} \frac{1 + g^*_{c,1}}{1 + g^*_{a,n}}.
\label{x2_1}
\eeq  
Comparing the expressions for $\chi$ in terms of ${\bar g}$ and $g^*$, 
 we read off the components of $g^*$ as
\beq
g^*_{a,n} = g_{a,n} + {\bar g}_{a,n} \left (1 + g_{a,n} \right).
\label{y_11}
\eeq 
 For all partial components different from the $n=2$ charge one,
 the regular contribution $g_{a,n}$ is absent, and 
$g^*_{a,n} =  {\bar g}_{a,n}$. 
The divergence of ${\bar g}_{a,n}$  implies that {\it  all components of 
 the actual Landau function $g^*$ different from the 
 nematic one diverge at a Pomeranchuk instability}. 
The $n=2$ component behaves as 
\bea
&&
g^*_{c,2} =  \left(1 + {\bar g}_{c,2}\right) \left (1 + g_{c,2} \right)-1 \nonumber\\
&\approx&
\lambda(1 +g_{c,2})-1=\frac{(1+g_{c,2})^{1/2}}{2a\kf}-1.
\label{y_11a}
\eea 
The nematic component of the actual Landau function 
approaches $-1$ at a QCP, where $g_{c,2} =-1$.
We show the behavior of ${\bar g}_{a,n}$ and $g^*_{a,n}$ in Fig. \ref{fig:sigma_1}.

Note in passing that, although  $\chi_{c,2}
$ diverges as $1/(1 + g_{c,2})$ 
in both the ordinary and critical FL regimes, this divergence has different origins in the two regimes. In an ordinary FL (where $m^*/m\sim 1$), the divergence of $\chi_{c,2}
$ is entirely due to that of the  
$d-$wave charge  
\lq\lq g-factor\rq\rq $1/(1+g_{c,2})$,
 while in the critical FL the divergence 
 is  equally \lq\lq shared\rq\rq\ between the effective mass
 and the
 \lq\lq g-factor\rq\rq\/, 
 each contributing a factor of $1/(1 + g_{c,2})^{1/2}$.  
  The analogous \lq\lq sharing\rq\rq\/ holds for  the spin susceptibility near a ferromagnetic instability.

\section{Equivalence of the one-loop and exact results for the self-energy}
\label{sec:4a}
In effective low-energy theories of QCPs, e.g., in the spin-fermion model,
the effective interaction is $D(Q)=g^2_{\mathrm{FB}}\chi(Q)$, where $\chi(Q)$  is the susceptibility of the divergent order parameter and $g_{\mathrm{FB}}$
is the fermion-boson coupling. The self-energy is obtained via a loop-wise expansion in $D$, which is usually truncated at the one-loop order (cf. Fig.~\ref{fig:sigma}{\it a})
\beq
\Sigma_{1L} (K) = \int D (K-P) G (P) \frac{d^3 P}{(2\pi)^3}.
\label{wed_11}
\eeq 
Such a formula is
  used in the Eliashberg and FLEX theories (with 
the bare or full $G (P)$, respectively\cite{regularize,FLEX}). 

On the other hand, we showed in the previous Section that the linear in $\omega$ and $\epsilon_k$ parts of the $\Sigma$ can be found from the Pitaevski-Landau
identities using a fully renormalized vertex. Now we can ask the following question: how does the one-loop result for $\Sigma$ correspond to that obtained from the Pitaevski-Landau identities? 
In this Section, we show
 that  Eq.~(\ref{wed_11}) is asymptotically exact in the critical FL regime
 (with corrections small in 
 $1/a\kf$ and $\omega/\omega_{\mathrm{FL}}$,
  if the effective interaction $D(Q)$ 
  is identified with 
$Z \Gamma^\Omega$ (more accurately, with  $\sum_\beta 
Z \Gamma^\Omega_{\alpha\beta,\alpha\beta}$). 

To prove this assertion, we  
  substitute $\sum_\beta Z \Gamma^\Omega_{\alpha\beta,\alpha\beta}$ for $D$ 
into  Eq.~(\ref{wed_11}), use 
 Eq.~(\ref{wed_66}) for $\Gamma^\Omega$, replace $G$ by its coherent part, and take
the limit of small  $\omega$ and $\epsilon_k$.
The effective interaction can be decomposed into static and dynamic parts.
 The static part gives an \lq\lq anomalous\rq\rq\/ $\epsilon_k$ term in $\Sigma_{1L}$, which comes from the immediate vicinity of the FS. The dynamic part gives a  regular $\omega-\epsilon_k$ term, which comes from the entire phase space. 
Explicitly, $\Sigma_{1L} (K) = \Sigma^{\mathrm{an}}_{1L} (K) + \Sigma^{\mathrm{reg}}_{1L} (K)$ 
\begin{subequations}
\begin{widetext} 
\bea
\Sigma^{\mathrm{an}}_{1L} (K) &=& \epsilon_k \frac{Z^2 m}{4 \pi^2}
 \sum_\beta \int \Gamma^\Omega_{\alpha \beta, \alpha \beta}
 (\theta) {\cos \theta} d \theta\label{wed_12a}\\
 \Sigma^{\mathrm{reg}}_{1L} (K) &=& \left(\omega-\epsilon_k \frac{m}{m^*}\right) (-i) \sum_\beta \int \Gamma^\Omega_{\alpha \beta, \alpha \beta} (K_F, P) \left[G^2 (P)\right]_\Omega \frac{ d^3 P}{(2\pi)^3}.
\label{wed_12b}
\eea
\end{widetext}
\end{subequations}
We see immediately that the $\omega$ terms in
Eqs.~(\ref{wed_2a}) and (\ref{wed_12a},\ref{wed_12b}) are the same,
whereas the $\epsilon_k$ terms are the same if 
\bwt
\beq
\frac{Z\left(Z-1\right)m}{4\pi^2}\sum_\beta \int \Gamma^\Omega_{\alpha \beta, \alpha \beta}
 (\theta) {\cos \theta} d \theta+\left(\frac{m}{m^*}-1\right)i\sum_\beta \int \Gamma^\Omega_{\alpha \beta, \alpha \beta} (K_F, P) \left[G^2 (P)\right]_\Omega \frac{ d^3 P}{(2\pi)^3}=0.
 \label{ident}
\eeq 
\ewt
Expressing the integrals of the vertices in Eq.~(\ref{ident}) via $m^*$ and $Z$ using Eqs.~(\ref{wed_4a}) and (\ref{wed_4b}),
 we see that Eq. (\ref{ident}) reduces to an identity
\beq
 - \left(\frac{m}{m^*}-1\right) \left[\frac{1}{Z}-1\right]  + (Z-1) \left[\frac{1}{Z} \left(1 - \frac{m}{m^*}\right)\right] =0
\label{wed_15}
\eeq
 We thus see that both $\omega$ and $\epsilon_k$ terms in 
the one-loop self-energy  coincide
with the exact expressions obtained using the Pitaevski-Landau identities, if
 $Z \Gamma^\Omega$ is identified 
 with the effective interaction.
This equivalence indeed
holds only as long as $m^*/m$ and $Z$ are energy independent and 
only a coherent part of $G(Q)$ is relevant,
which is the case  
for the critical FL regime of a QCP.

In Appendix~\ref{sec:subtle}, we show  explicitly how the self-energy is reproduced by the diagrammatic exapnsion for $\Gamma^\Omega$.

\section{Conclusions}
\label{sec:5}

In this paper, we analyzed properties of the Fermi liquid 
 near a quantum phase transition, using a simple model of the nematic n=2 
 charge  Pomeranchuk instability as an example.
 Our main result is that, near a phase transition, the system enters into a 
new critical
 FL regime, in which 
  all spin Landau components and all charge components 
with $n \neq 2$
increase and eventually diverge at the critical point.
 This behavior is the consequence of a singular momentum dependence 
of the Landau function 
in a critical FL. 
 Therefore, 
 a common assumption that all
Landau components,
other than the one corresponding to the critical channel
($g_{c,2}$ for the nematic charge instability),
are featureless near a 
 transition is  
 incorrect in $D \leq 3$.
 The divergence of the Landau components, including the one controlling renormalization of the effective mass, has no
consequences for susceptibilities channels because the divergent effective mass cancels out with the divergent effective
\lq\lq $g$-factor\rq\rq\/. 
 
To prove these statements, we derived the Landau function for a critical Fermi liquid, related to the vertex $\Gamma^\Omega$ via
$g = 2\nu  Z^2 (m^*/m) \Gamma^\Omega$, where $\nu$ is the density of states, $Z$ is the quasiparticle residue, $m^*$ is the effective mass, and
$\Gamma^\Omega$ is defined in the limit of zero momentum transfer and vanishingly small energy transfer.
Our starting point is a model of 2D fermions with a 
$d-$wave interaction, which we assumed to be of sufficiently long range, such that $a\kf\gg 1$, where $a$ is the effective radius of the interaction.  
 We computed the Fermi-liquid vertex $\Gamma^\Omega$  in two stages. First, we considered only those diagrams for $\Gamma^\Omega$ 
 that do not contain soft particle-hole bubbles
 with zero momentum and vanishing frequency.
 For $a\kf\gg 1$, diagrams of that type can be summed up 
in the RPA.  
 The RPA vertex $\Gamma^{\Omega, \mathrm{RPA}}$  contains a component which diverges at the critical point and can be interpreted as 
 arising from the exchange by soft collective excitations 
 in the $n=2$ charge channel. 
Next, we included the diagrams with soft particle-hole bubbles.
In an ordinary FL, the full interaction is approximated by its static term, and such diagrams vanish due to particle-number conservation. 
 In a critical FL, however, 
 the effective interaction is 
 dynamic, and the diagrams with soft particle-hole bubbles are 
finite.  We showed that 
  the relevant diagrams of 
 that  kind
  form a ladder series. Summing up this series, we found
 that non-RPA diagrams with soft bubbles renormalize the RPA vertex by a constant factor, i.e.,  
 the full vertex $\Gamma^\Omega$ is  equal to $\Gamma^{\Omega, \mathrm{RPA}}/Z_{\Gamma}$, where $Z_{\Gamma}$ is
 a function of $Z, m^*/m$, and of the dimensionless coupling constant 
$\lambda 
=1/(2a\kf \sqrt{1+g_{c,2}})$.
 Vertex corrections to ladder diagrams were shown to be 
  small 
 as $1/a\kf$.    Having found $\Gamma^\Omega$, we employed the Pitaevski-Landau identities
  for the derivatives of the self-energy in terms of $\Gamma^\Omega$
  to obtain
  coupled  equations for $Z$ and $m^*/m$.
This gave us a closed set of equations for $Z_\Gamma$, $Z$, and $m^*/m$ with $\lambda$ as a parameter. 
Solutions 
 of these equations 
allows one to follow 
 the evolution 
 from an ordinary FL behavior at $\lambda <1$, where 
$m^*/m$,  $Z$ and $Z_\Gamma$ are all close to 1, to a critical FL behavior at $\lambda\gg 1$, where
$m^*/m = 1/Z \approx \lambda$ and $Z_\Gamma \sim Z$.
We also showed that the self-energy of a critical FL is \lq\lq local\rq\rq\/, i.e.,  
$\Sigma (k, \omega) \approx \Sigma (\omega)$, and that $m^*$ 
diverges only at $\lambda=\infty$ rather than at finite $\lambda$. The latter result implies that no
preemptive transitions occur before the nematic one is reached. 
 
 Finally, we showed that the exact self-energy (as given by the Pitaevski-Landau identities)
 is asymptotically close to the  
one-loop expression $\Sigma (K) \propto \int G(Q) D(K-Q)$
 if one identifies the effective interaction 
  $D$ with $Z \Gamma^\Omega=\Gamma^{\Omega, \mathrm{RPA}}$.  We have shown explicitly that in the limit of $\omega,\epsilon_k\to 0$, corrections to the one-loop result are  small in powers of 
 $1/a\kf$ and $\omega/\omega_{\mathrm{FL}}$, where $\omega_{\mathrm{FL}}$ is an upper bondary for the FL behavior.

\section{Acknowledgment}

We acknowledge helpful discussions with0
 C. Castellani, L. Dell'Anna, M. Garst, I. Eremin, E. Fradkin,  H.-Y. Kee, 
Eun-Ah Kim,  Y.-B. Kim, S. Kivelson, M. Lawler, W. Metzner, L.P. Pitaevski, 
A. Rosch, T. Senthil, A. Varlamov,  T. Vojta, P. W{\"o}lfle, and C. Wu.
 We are particularly thankful to M. Garst and P. W{\"o}lfle for sharing their unpublished results with us. This work was supported by  NSF-DMR-0908029 (D. L. M.) and NSF-DMR-0906953
 (A. V. Ch.) Partial support
 from MPI PKS (Dresden) [D. L. M. and A. V. Ch.] and Basel QC2 visitor program [D. L. M] is gratefully acknowledged. 

\appendix

\appendix

\section{Diagrammatic series for $\Gamma^\Omega$ beyond RPA}
\label{app:b}

\begin{figure}[tbp]
\centering
\includegraphics[width=1.0\columnwidth]{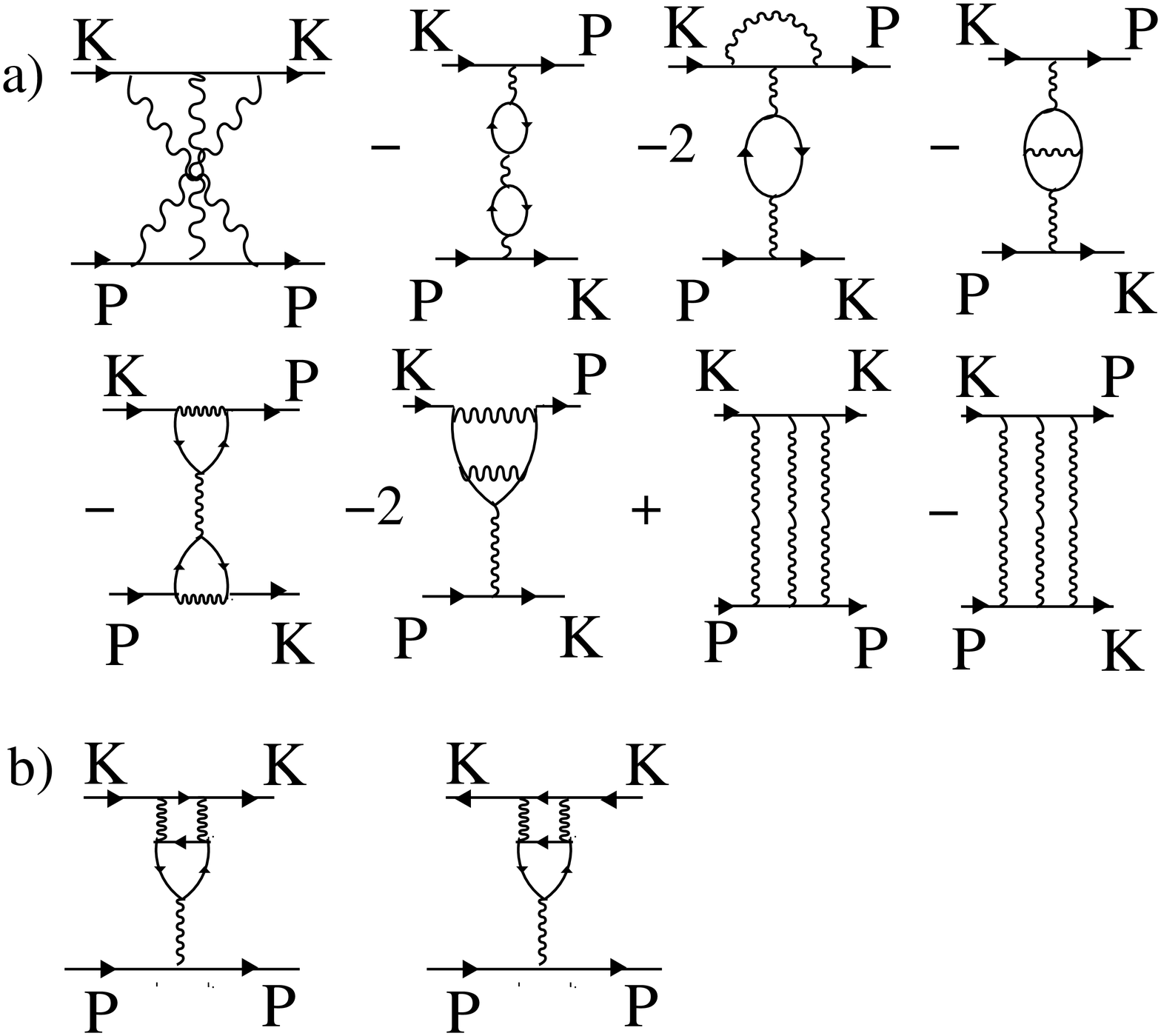}
\caption{a) Third-order diagrams for $\Gos$ that contain only the particle-particle and particle-hole bubbles at combinations of external momenta $K\pm P$.
b) Examples of the third-order diagrams that contain convolutions of particle-particle and particle-hole bubbles with Green's functions over the internal momenta.}  
\label{fig:gamma3}
\end{figure}
In the main part of the text, we adopted a model case of the long-range interaction in the the $d$-wave channel to justify the RPA. Such a model, although a necessary pre-requisite for an RPA-like treatment in any channel, is hardly realistic and one is naturally led to wonder as to what extent the main features of the effective interaction, Eq. (\ref{t_10a}), survive beyond the RPA level.
In this Appendix,  we address this issue by evaluating 
$\Gamma^\Omega$ in a direct order-by-order perturbation theory for a short-range, Hubbard-like interaction ($U=$const).  
Notice that this issue is different from renormalization of the RPA interaction via ladder series in the dynamic interaction, considered in Sec.~\ref{sec:3_1} of the main text.
Here, we discuss  the full $\Gamma^\Omega$ without contributions
from the soft partcle-hole bubbles, which is the input for the calculations in 
Sec.~\ref{sec:3_1}. The issue is whether this full $\Gamma^{\Omega}$ has the same structure as the 
RPA result, Eq. (\ref{t_10a}), i.e., whether it can be interpreted as the exchange by soft collective excitations. 

The issue of validity of the RPA  is not specific to a charge nematic QCP. To avoid unnecessary complications associated with the $d-$wave factors in the vertices, we consider here a simpler case of an
instability in the $n=0$ channel (charge instability for $U <0$ or spin instability for $U >0$). 

Diagrams for $\Gamma^\Omega$ to second order are shown in Fig. ~\ref{fig:gammaomega} {\it a}). Explicitly,
\bwt
\bea
&&\Gos (K, P) = -  \frac{U}{2} \left(1 - U \Pi_{ph} (K-P) + 
U \Pi_{pp} (K+P)\right) \delta_{\alpha\delta}\delta_{\beta\gamma} \nonumber \\
&& + \frac{U}{2} \left(1 + U \Pi_{ph} (K-P) + U \Pi_{pp} (K+P)\right) {\bf \sigma}_{\alpha\delta}{\bf \sigma}_{\beta\gamma},
\label{b_1}
\eea
\ewt
where $\Pi_{ph} (K) = - \int (d^3L/(2\pi)^3) G_L G_{L+K}$ and $\Pi_{pp} (K) = - \int (d^3L/(2\pi)^3) G_L G_{-L +K}$ are particle-hole and particle-particle bubbles.  
Notice that, to this order, all contributions to $\Gos$ contain either $\Pi_{ph} (K-P)$ or $\Pi_{pp} (K+P)$ but no bubbles with other momenta.

The new physics emerges at the third order of the interaction. The third-order diagrams can be divided into two classes. Diagrams from the first class, shown in Fig.~\ref{fig:gamma3}){\it a} still
 contain either $\Pi_{ph} (K-P)$ or $\Pi_{pp} (K+P)$. Diagrams from the second class contain convolutions of $\Pi_{ph}$ and $\Pi_{pp}$. 

Adding the contributions of the third order diagrams from the first class to Eq.~(\ref{b_1}), we obtain
\bwt
\bea
&&\Gos (K, P) = -  \frac{U}{2} \left(1 - U \left(\Pi_{ph} (K-P) - \Pi_{pp} (K+P)\right) + U^2 \left(\Pi^2_{ph} (K-P) + \Pi^2_{pp} (K+P)\right)\right)
 \delta_{\alpha\delta}\delta_{\beta\gamma} \nonumber \\
&& + \frac{U}{2} \left(1 + U \left(\Pi_{ph} (K-P) + U \Pi_{pp} (K+P)\right)
 +  U^2 \left(\Pi^2_{ph} (K-P) + \Pi^2_{pp} (K+P)\right)\right) 
 {\vec \sigma}_{\alpha\delta}\cdot{\vec \sigma}_{\beta\gamma}.
\label{b_2}
\eea  
\ewt
This result for $\Gos (K,P)$ can be re-written as an expansion of the  following formula 
\bwt
\bea
\Gos (K, P) &=& -  \frac{U}{2} \left[\frac{1}{1 - U \Pi_{pp} (K+P)} + 
\frac{1}{1 + U \Pi_{ph} (-P)}-1\right] 
\delta_{\alpha\delta}\delta_{\beta\gamma}  \nn\\&&+ \frac{U}{2}
\left[\frac{1}{1 - U \Pi_{pp} (K+P)} 
\frac{1}{1 - U \Pi_{ph} (K-P)}-1\right] 
 {\vec \sigma}_{\alpha\delta}\cdot{\vec \sigma}_{\beta\gamma}.
\label{b_3}
\eea
\ewt
Although the last formula is valid only to order $U^3$, 
it is very likely that higher-order diagrams for $\Gamma^\Omega$ 
of the same type, i.e., containing only $\Pi_{ph} (K-P)$ or $\Pi_{pp} (K+P)$,  are described by this expression. We see from Eq.~(\ref{b_3}) that the structure of the vertex is virtually the same as in the RPA in a sense that $\Gos$ contains three separate geometric series,  describing exchange by charge,
spin, and pairing fluctuations.
The three separate contributions to $\Gos$ 
diverge near a corresponding QCP, e.g., near $U \Pi_{ph} (0) =-1$ for a charge QCP and $U \Pi_{ph} (0) =1$ for a  spin QCP
 (the former can occur if pairing instability is suppressed by, e.g., magnetic field).
 We  emphasize that, at this level, the particle-particle and particle-hole channels 
 entirely decouple, i.e., the interaction in the pairing channel does not affect the structure of the effective interaction mediated by soft 
 collective excitations in the particle-hole channel.

The situation is changed by diagrams from the second class. There are 24 topologically distinct third-order diagrams which contains momentum integrals integrals of the polarization bubbles, and we refrain from presenting all of them. As an example, two of the 24 diagrams are shown in Fig.~\ref{fig:gamma3}
{\it b}).   We computed analytically all 24 diagrams. 
There are numerous cancellations between the diagrams,
 and the final result is rather compact: 
\bwt
\bea  
\G (K,P) &=& \Gos (K,P)  - U^3 \left[\int \frac{d^3L}{(2\pi)^3} \left[\Pi_{ph} (L) \left(2G_{L+K}G_{L+P}-G_{L+K}G_{P-L}\right) + \Pi_{pp} (L) G_{L-P}G_{L-K}\right] \delta_{\alpha\delta}\delta_{\beta\gamma} \right] \nonumber \\
&& + U^3 \int \frac{d^3L}{(2\pi)^3} \left[\Pi_{ph} (L) G_{L+K}G_{P-L} + \Pi_{pp} (L) G_{L-P}G_{L-K}\right]
 {\vec \sigma}_{\alpha\delta}\cdot{\vec \sigma}_{\beta\gamma},
\label{b_4}
\eea
\ewt
where 
$\Gos$ is given by Eq.~(\ref{b_3}). 

We see from Eq.~(\ref{b_4}) that there are 
 cross terms which involve both
 particle-hole and particle-particle bubbles, i.e., the terms with $\Pi_{ph} (L) G_{L+K}G_{P-L}$ and $\Pi_{pp} (L) G_{L-P}G_{L-K}$).  The presence of such terms implies that the particle-particle and particle-hole channels do couple beginning from the third order in $U$. This coupling should affect 
 the interaction mediated by near-critical
 charge- and spin-density fluctuations.

Note that ${\bar \Gamma}$ contains a term
 which involves only particle-hole bubbles[($\Pi_{ph} (X) G_{X+K}G_{X+P}$].
To understand qualitatively the effect of this term, we neglect momentarily other terms and approximate $\Pi_{ph}$ by a constant ( we recall that
 a static bubble $\Pi_{ph} ({\bf q},0)$ is a constant  for $q< 2\kf$ in 2D). After simple manipulations, we then obtain   
\bea
&&\G (k, p) = - U \left[\frac{1}{1 + U^2 \Pi^2_{ph}} 
- \frac{1}{2}\frac{1}{1 - U \Pi_{ph}}\right] 
\delta_{\alpha\delta}\delta_{\beta\gamma} \nonumber \\
&& + \frac{U}{2}
\frac{1}{1 - U \Pi_{ph} (K-P)} 
 {\vec \sigma}_{\alpha\delta}\cdot{\vec \sigma}_{\beta\gamma}.
\label{b_5}
\eea  
We see that 
 the 
 term 
 $1/(1 + U \Pi_{ph})$
 describing  the interaction
 via  soft bosons in the charge channel
 is no longer there. Although 
 disappearance of this term well may be an artifact of the approximation [Eq.~(\ref{b_5}) is, strictly speaking valid only to order $U^3$], this is still clearly a warning sign for the
 whole approach, as it shows that additional terms with particle-hole bubbles, not included into RPA-type analysis may be relevant. 
Note also that there is a ``leakage'' of the $1/(1-U \Pi_{ph})$ term  from the spin channel into the charge channel, i.e  interaction mediated by a soft boson in the spin channel induces the same interaction in the charge channel. This effect was considered in detail in Ref.~\onlinecite{spin_we}.

\section{
Alternative averaging procedures for $\Gamma^\Omega$}
\label{app:a}

For completeness, we present here the results for $Z_\Gamma$, $Z$, and $m^*/m$
 obtained 
   within 
   alternative approaches for 
   summing up the diagrammatic series (\ref{gn}) for $\Gamma^\Omega$.  We remind that the uncertainty is related to the presence of the factors 
$d_\bk d_\bp (d^{2(n-1)}_\bk + d^{2(n-1)}_\bp)$ in the diagrammatic series for $\Gamma^{\Omega}_{\alpha \beta, \gamma \delta} ({\bf k}, {\bf p})$. In the main part of text, we used the approximation in which
 the angular dependence of $\Gos$ was averaged over the FS. An alternative is to keep only the $d_\bk d_\bp$ term at each order and neglect other (non-$d-$wave) terms. A simple exercise 
 in trigonometry shows that this amounts to replacing 
\beq
d_\bk d_\bp (d^{2(n-1)}_\bk + d^{2(n-1)}_\bp) \rightarrow (2n-1) d_\bk d_\bp
\label{ch1}
\eeq
at each order of the perturbation theory.
Summing up 
 the series for  $\Gamma^{\Omega \{n\}}_{\alpha \beta, \gamma \delta} ({\bf k}, {\bf p})$, we 
 find that Eq.~(\ref{gfull}) is still valid,
i.e., $\Gos= \GRPA/Z_{\Gamma}$, where $Z_\Gamma$ is a constant,  but now the expression for $Z_{\Gamma}$ is different:
\bea
\frac{1}{Z_{\Gamma}}&=&1 + \sum_{n=1}^{\infty}\left(\frac{Z^2m^*}{2m}\lambda\right)^{n} \left(n+\frac{1}{2}\right)\nn\\
&&=
\frac{2 - \frac{Z^2m^* \lambda}{2m}\left(1-\frac{Z^2m^* \lambda}{2m}\right)}{2\left(1 - \frac{Z^2m^*\lambda}{2m}\right)^2}.
\label{zgamma1}
\eea
Using this form and 
 isotropic FL formulas for $m^*/m$ and $Z$, we
 obtain after simple algebra that in the critical FL regime, where $\lambda$ is large, 
  $m^*/m = 1/Z = \lambda/2$  and $Z_{\Gamma} \approx 1/2\lambda \approx Z/2$. 

We see that, as in  Eq.~(\ref{mz}),  $Z_{\Gamma} \big\vert_{\lambda\gg 1} 
\sim Z$, the only difference between  (\ref{mz}) and the present case 
 being the numerical prefactor. This proves our point that both procedures 
 for summing up the ladder series for 
$\Gamma^\Omega$ yield physically equivalent results.

We also obtain very similar results 
 by neglecting the 
$d-$wave structure of the intermediate vertices in the ladder series, 
 i.e., 
replacing $(d^{2(n-1)}_\bk + d^{2(n-1)}_\bp)/2$ by $(1/2)^{n-1}$. In this situation, we
 obtain $1/Z_\Gamma = 1/(1 -\frac{Z^2m^* \lambda}{4m})$. 
 Combining this with the
 equations for $Z$ and $m^*/m$ and solving the full set, we  obtain
  $m^*/m = 1/Z =\lambda/4$, and $Z_{\Gamma} \big\vert_{\lambda\gg 1}  = Z$ for $\lambda\gg 1$.\\

\section{Loop expansion for $\Sigma$}
\label{sec:subtle}

\subsubsection{Relation between vertex renormalization and single-particle residue}
 We begin by discussing a  subtle point in the relation 
between the effective fermion-boson interaction 
 $D=g_{\mathrm{FB}}^2\chi$
  and  $Z \Gamma^\Omega$. We found that in the critical 
FL regime, the vertex renormalization constant 
$Z_\Gamma$ is related to the single-particle residue $Z$ as $Z_{\Gamma}=C Z$, where $C\sim 1$ depends on which of the approximate averaging procedures, discussed in Sec.~\ref{sec:3_1} and Appendix \ref{app:a}, is employed. One of such procedures yields $C=1$, whereas the other two yield $C \neq 1$. 
 The uncertainty is related to the fact that we used anisotropic  $d-$wave form of the interaction between fermions, yet approximated the quasiparticle residue $Z$ and the effective mass $m^*$ by constants 
 independent 
 of the position on the FS. 
Regardless of a particular value of $C$,   proportionality between $Z_{\Gamma}$ and $Z$ implies that $Z \Gamma^\Omega = (Z/Z_\Gamma) \Gamma^{\Omega, \mathrm{RPA}}$ is proportional to $\Gamma^{\Omega, \mathrm{RPA}}$.  It is then 
 instructive to verify whether the exact Pitaevski-Landau formula for $\Sigma$ is reproduced if 
 the diagrammatic expansion for the vertex $\Gamma^\Omega$ in terms of $\Gamma^{\Omega, \mathrm{RPA}}$ is transformed into an expansion for the self-energy  by contracting
 a pair of external legs of a vertex. The first term in the series is $\Gamma^{\Omega, \mathrm{RPA}}$, and hence the ``bare'' self-energy is the one with $\Gamma^{\Omega, \mathrm{RPA}}$ instead of $D$.  This would be the final result if $Z$ and $Z_\Gamma$ were
 equal in the critical FL (in that case, $\chi = Z \Gamma^\Omega = (Z/Z_\Gamma) \Gamma^{\Omega, \mathrm{RPA}} = \Gamma^{\Omega,\mathrm{RPA}}$). 

Whether $Z_\Gamma$ is equal to $Z$  or just proportional to $Z$ does not affect our main result that all Landau parameters diverge upon approaching a 
 nematic QCP. It turns out, however, that only an exact equality $Z_\Gamma = Z$ is consistent with the 
  loop-wise expansion for the self-energy.
 This is not
 surprising because an equality $Z_\Gamma = Z$ is the result of 
 a particular averaging 
  procedure in which
 the $d$-wave vertices arising from intermediate states
  were replaced by constants  when evaluating the ladder series for $\Gamma^\Omega$.   This is 
   consistent with 
   replacing $Z$ and $m^*$ by angle-independent constants. In the other two
  procedures, we either averaged the angle-dependent interaction over the FS at each order or 
extracted the  $d-$wave component from the interaction.  In both cases, there are extra combinatorial factors associated with the  $d-$wave nature of the interaction.

To see that only an equality $D = \Gamma^{\Omega,\mathrm{RPA}}$ is consistent with 
 diagrammatics, we consider explicitly diagrams for $\Sigma$.
To second-loop order, there are two diagrams, {\it b} and {\it c} in Fig.~\ref{fig:sigma}.  Diagram {\it b} is a part of the self-consistent renormalization of the Green's function in the one-loop diagram. 
 Diagram {\it c} is a vertex correction to the one-loop self-energy.
 In two subsequent Sections, we will compute these two diagrams and show that they are indeed small, at least as along as $a\kf\gg 1$. 

\subsubsection{Green's function renormalization in the one-loop self-energy}

\begin{figure}[tbp]
\centering
\includegraphics[width=0.9\columnwidth]{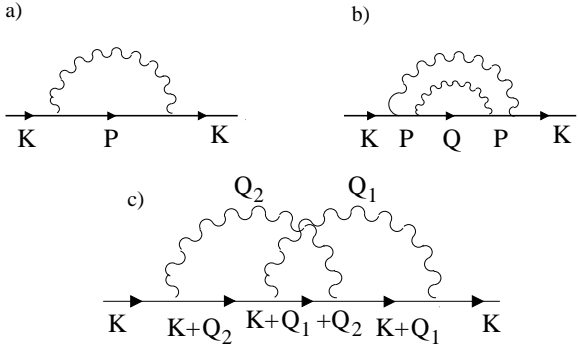}
\caption{Diagrams for the self-energy at the one-loop [{\it a})]  and two-loop orders [{\it b}) and {\it c})]. The  wavy line is the RPA interaction, 
Eq.~(\protect\ref{t_10a}). }
\label{fig:sigma}
\end{figure}

We start with diagram {\it b} in Fig.~\ref{fig:sigma}:
 \beq
\Sigma^{\{b\}}_{2L} \sim \int d^3 P d^3 Q \left[G^2 (P)\right]_\Omega G(Q) D(K-P)
 D (Q-P). 
\label{wed_18}
\eeq
Integrating over $Q$ first, we obtain
\beq
\Sigma^{\{b\}}_{2L} \sim \int d^3 P \left[G^2 (P)\right]_\omega  D (K-P) 
\Sigma^{\{1\}} (P),
\label{wed_18a}
\eeq 
where $\Sigma_{1L}(P)$ is one loop-self-energy with $D = \Gamma^{\Omega, \mathrm{RPA}}$.
The leading term in $\Sigma_{1L} (P)$ is 
 $\lambda \omega$. This term, however, vanishes
upon integration over $P$. The most straightforward way to see this is to perform momentum integration first. Replacing $\int d^2 p$ in Eq.~(\ref{wed_18a}) by $\int d \epsilon_p$ and integrating over $\epsilon_p$ from $-\Lambda$ to $\Lambda$, where $\Lambda \sim \ef$, we find after simple algebra that, because of the double pole in $G^2 (P)$, the integral comes either from vanishingly small $\epsilon_p$
or from $|\epsilon_p| \sim \Lambda$. 
The high-energy contribution to $\Sigma^{\{2\}}$ is of order 
\bea
&&\frac{Z^2 m^*}{m} \lambda  \int d\omega_p \omega_p \int^\Lambda_{-\Lambda} \frac{d\epsilon^*_p}{(\omega_p - \epsilon^*_p)^2} \nonumber \\
&&\sim \int_0^\Omega d \omega_p \omega_p \frac{2\Lambda}{\omega^2_p - \Lambda^2} \sim  \frac{\omega^2}{\Lambda} \ll \Sigma_{1L}.
\eea
The contribution from vanishingly small $\epsilon_p$
\beq
{\bar I} = \frac{Z^2 m^*}{m} \lambda^2 \int d \omega_p \omega_p \int \frac{d \epsilon^*_p}{(\omega_p -\epsilon^*_p + i \delta ~\mathrm {sgn} \epsilon_p
)^2}
\label{wed_20a}
\eeq
comes from the branch cut at $\epsilon_p=0$, caused by
$\mathrm {sgn} \epsilon_p$ term in Eq.~(\ref{wed_20a}). Evaluating the integral, we find that it vanishes in the limit of $\delta\to 0$:
\beq 
{\bar I} \sim i \frac{Z^2 m^*}{m} \lambda^2  \int_0^\Omega d \omega_q \omega_q \frac{\delta}{\omega^2_p + \delta^2} \rightarrow 0.
\label{wed_20}
\eeq
A non-zero contribution to $\Sigma^{\{b\}}_{2L}$ comes only from the $\epsilon_k$ dependent term in $\Sigma_{1L}$, given by Eq. ~(\ref{wed_17}):
\begin{widetext}
\beq
\Sigma^{\{b\}}_{2L}\sim \frac{Z^2 \Delta m^*}{m}   \int d \omega_p d^2 p \frac{\epsilon^*_p}{(\omega - \epsilon^*_p + i \delta_\epsilon)^2} D (K-P) \sim \frac{Z \Gamma_{\triangle} m^*}{m}  \int d \omega_p d^2 p G(P)  D (K-P) \sim \Gamma_{\triangle} \Sigma_{1L}. 
\label{z_5}
\eeq
\end{widetext}
As we have shown in Sec.~\ref{sec:fullsigma}, 
$\Gamma_{\triangle} \sim
1/a\kf\ll 1$  i.e., $\Sigma^{\{b\}}_{2L}$ is indeed small compared to 
$\Sigma_{1L}$.   
The same conclusion holds for higher-order diagrams of the same type.
\subsubsection{Vertex corrections to the one-loop self-energy.} 
\label{sec:sigma_vertex}

Next, we consider the vertex correction to the one-loop self-energy given by diagram {\it c} in Fig.~\ref{fig:sigma}. Integrating over $P$, we obtain
\beq
\Sigma^{\{c\}}_{2L}(K)=-\int\frac{d^3Q}{(2\pi)^3}\Gamma_{\triangle}(Q;K)D(Q)G(K+Q).
\label{sigmac}
\eeq
In Sec.~\ref{sec:vertex}, we computed the three-leg vertex $\Gamma_{\triangle}(Q;K)$ in the limit of $\Omega/q\to 0$. In this limit, 
$\Gamma_{\triangle}\sim  1/a\kf\ll 1$ 
coincides with the effective Migdal parameter of our theory.
 To calculate the self-energy, however, we need  to know $\Gamma_{\triangle}(Q;K)$ at arbitrary $\Omega/q$. When evaluating the dynamic part of the vertex, it is important to account for finite curvature of the Fermi surface, i.e., to keep a quadratic term in the expansion of the dispersion
\beq
\epsilon^*_{\bk_{\mathrm{F}}+\bq} = \vf^*q_{\parallel} + \frac{q^2_{\perp}}{2m^*},
\label{disp}
\eeq
where $q_{\parallel}$ and $q_{\perp}$ are the components of $\bq$ along and transverse to ${\bf k}_{\mathrm{F}}$, correspondingly. 
In general, $\Gamma_{\triangle}(Q;K)$ is an involved function of its arguments; however, a useful estimate can be written as
 \beq
\Gamma_{\triangle} (\bq, \Omega;\bk_{\mathrm{F}},\omega_k=0) = 
 \Gamma_{\triangle} +
 \lambda \frac{Z^2 m^*}{m} \frac{|\Omega|}{\max\{|\Omega|, 
|\epsilon^*_{\bk_{\mathrm{F}}+\bq}|\}},
\label{zz_3}
\eeq
 where the first term is the vertex in the limit $\Omega/q\to 0$.
 As it is always the case in Migdal-Eliashberg--type theories, the vertex depends crucially on the ratio of energies entering the dynamic term.
Substituting the dynamic term into the self-energy, one can readily see that 
$\Omega$ is of order of the external frequency $\omega_k$, $q_{\parallel}$ is of order $\omega/\vf^*$, and $q_{\perp}$ is of order of the inverse correlation length
 $\xi^{-1}=(1 + g_{c,2})^{1/2}/a$,
 so that a typical value of $|\epsilon^*_{\bk_{\mathrm{F}}+\bq}|$ is the largest of the two energies--
 $|\omega_k|$ and $\omega_{\mathrm{FL}}$,
where
\beq
\omega_{\mathrm{FL}} = \frac{1}{m^*\xi^2}=\frac{\ef}{a\kf } \left(1+ g_{c,2}\right)^{3/2}.
\label{ch_6}
\eeq         
Therefore, the vertex can be estimated as  
\beq
\Gamma_{\triangle} (\bq, \Omega;\bk_{\mathrm{F}},\omega_k=0) \sim  
 \Gamma_{\triangle} + \frac{|\Omega|}{\max\{|\Omega|,\omega_{\mathrm{FL}}\}}.
\label{zz_44}
\eeq
The physical meaning of $\omega_{\mathrm{FL}}$ is that it is an upper boundary for a FL behavior near a QCP. 
For energies higher
 than  
$\omega_{\mathrm{FL}}$, Landau damping becomes the dominant term  in $\Gamma^{\Omega, \mathrm{RPA}}$ and, consequently, $Z$ and $m^*$ become energy dependent.
The FL theory is therefore valid only at energies
 smaller than $\omega_{\mathrm{FL}}$. We see from Eq.~(\ref{zz_44}) that 
 vertex corrections are irrelevant for 
$\Gamma_\triangle \ll 1$ and $|\omega|\ll \omega_{\mathrm{FL}}$. 

 Notice that the smallness of dynamic term in $\Gamma_\triangle$ is due to the presence of the 
quadratic term $q_{\perp}^2/(2m^*)$ in the fermionic dispersion, Eq.~(\ref{disp}). 
This term reflects finite curvature of the FS in 
 larger than one dimensions. Without such a term, vertex correction would be of order one for any frequency.
Importance of the the FS curvature for vertex corrections in 2D 
has been discussed in Refs.~\onlinecite{aim,rech,CK06,tigran,sslee}.

To  verify this reasoning, we  computed diagram {\it c} in Fig.~\ref{fig:sigma} explicitly.  There are two contributions to the self-energy: the first one is linear in $\omega$ and gives a correction to mass renormalization; the second one is quadratic in $\omega$ and gives a correction to damping. Although the $\omega^2$ contribution is formally smaller than the linear one, the coefficient of the $\omega^2$ term diverges in the non-FL regime. 
The linear contribution can be extracted by keeping only
 $\Gamma_{\triangle}$ in (\ref{zz_3}) and substituting it into Eq. ~(\ref{sigmac}).
This gives
\beq
\Sigma^{\{c,1\}}_{2L} ({\bf k}_F, \omega)\sim 
\Gamma_{\triangle} \Sigma_{1L} \sim \Gamma_{\triangle} \lambda\omega.
\eeq
The $\omega^2$ contribution is obtained from the entire expression for the self-energy
\begin{widetext}
 \beq
\Sigma^{\{c,2\}}_{2L} ({\bf k}_F, \omega) \sim Z^3 \int d^2 q_1 d^2 q_2 d\Omega_1 d\Omega_2
 \frac{\Gamma^{\Omega, \mathrm{RPA}} (q_1, \Omega_1) \Gamma^{\Omega, \mathrm{RPA}} (q_2, \Omega_2)}
{\left[i\left(\omega + \Omega_1\right) - \epsilon^*_{\bk_{\mathrm{F}} +\bq_1}\right] \left[i\left(\omega + \Omega_1 + \Omega_2\right) - \epsilon^*_{ \bk_{\mathrm{F}} +\bq_1+\bq_2}\right]  \left[i\left(\omega + \Omega_2\right) - \epsilon^*_{\bk_{\mathrm{F}} +\bq_2}\right]},
\label{zz_55}
\eeq
\end{widetext}
where $\epsilon^*_{\bkf+\bq}$ is given by Eq.~(\ref{disp}). Integrating over $q_{||}$ first, we see that the region of integration over the internal frequencies, $\Omega_1$ and $\Omega_2$, is bounded by external $\omega$. This immediately shows that the double integral $\int\int d\Omega_1d\Omega_2$ contributes a factor of $\omega^2$.  Rescaling variables as $x=q_{1\perp}\xi$ and $y=q_{2\perp}\xi$, we obtain
\beq
\Sigma^{\{2,c\}}_{2L} ({\bf k}_F, \omega) \sim i\lambda \frac{|\omega|^3}{\omega_{\mathrm{FL}}^2} \int_0^\infty \frac{d x}{1+x^2} \int_0^\infty \frac{d y}{1+y^2} 
\frac{1}{\beta^2 + x^2 y^2},
\label{zz_66}
\eeq
where $\beta=|\omega|/\omega_{\mathrm{FL}}$. For $\beta\ll 1$,  the double integral in Eq.~(\ref{zz_66}) behaves as $\ln|\beta|/|\beta|$. Collecting the two contributions, the final result for diagram {\it c} reads
\beq
\Sigma^{\{c\}}_{2L}\sim\left(
\frac{\vf/a}{\ef} +i\frac{\omega}{\omega_{\mathrm{FL}}}\right)\Sigma_{1L}.
\label{sigmac_2}\eeq
We see that  $\Sigma^{\{2\}} $ is indeed parametrically smaller than $\Sigma_{1L}$ in the FL regime, 
i.e.,  vertex corrections 
are  also irrelevant for the  fermionic self-energy.

To conclude this Section, we compare our results with the canonical Migdal-Eliashberg theory of the electron-phonon interaction. In that theory,  the vertex correction to the self-energy has a similar form, except for the denominators of both terms in Eq.~(\ref{sigmac_2}) contain the same energy scale: the Fermi energy. In this case, the second term is always smaller than the first one as long as 
 $\omega$
is smaller than the typical phonon frequency.
  In our case, however, the effective \lq\lq Fermi energies\rq\rq\/ in the static and dynamic terms are different and, what is most important, they behave differently as a QCP is approached:
whereas $\ef$ 
 remains finite, the upper boundary of the FL behavior, $\omega_{\mathrm{FL}}$, vanishes  as $1/\lambda^3$. In the non-FL regime, the one-loop self-energy behaves as $\Sigma_{1L}\sim\omega_0^{1/3}\omega^{2/3}$ for $\omega\ll\omega_0$, where $\omega_0=\ef/(a\kf)^4$. This behavior can be interpreted as resulting from the energy dependences of the mass renormalization coefficient $\lambda(\omega)\sim  (\omega_0/\omega)^{1/3}$ and of the quasiparticle residue $Z(\omega)\sim 1/\lambda(\omega)$.  Re-writing $\omega_{\mathrm{FL}}$ as $\omega_0/\lambda^3$ with $\omega$-dependent $\lambda$, we see that $\omega_{\mathrm{FL}}\sim \omega$. Therefore, the second term in Eq.~(\ref{sigmac_2}) is of order unity, 
 i.e.,  vertex corrections are, in general, 
 important in the non-FL region 
 of a QCP and also for the $\omega^2\ln \omega$ term in the FL regime.\cite{sslee_1}  Some  of the higher-loop diagrams 
 can be regularized by expanding the theory to the large $N$ case\cite{aim,rech,tigran}; however, there are still diagrams that are not small in the large $N$ approximation. \cite{sslee,sslee_1}


\begin{thebibliography}{99}
\bibitem{comment_RPA} This situation is to be contrasted with a more familiar RPA for the Coulomb potential, where an assumption of the long-range screened interaction, required for controlling the RPA series, necessarily implies that the interaction is also weak. In that case, there is no need to go beyond the RPA which simply gives a weak-coupling result. In our case, the range of the interaction and its magnitude are two independent parameters, and the  condition $a\kf \gg 1$ does not guarantee convergence of the series beyond the weak-coupling limit.
\end{thebibliography}
\end{document}